\documentclass[iicol,sn-mathphys-ay]{sn-jnl}

\usepackage{graphicx}%
\usepackage{multirow}%
\usepackage{amsmath,amssymb,amsfonts}%
\usepackage{amsthm}%
\usepackage{mathrsfs}%
\usepackage[title]{appendix}%
\usepackage{xcolor}%
\usepackage{textcomp}%
\usepackage{manyfoot}%
\usepackage{booktabs}%
\usepackage{algorithm}%
\usepackage{algorithmicx}%
\usepackage{algpseudocode}%
\usepackage{listings}%

\usepackage{microtype}
\usepackage{subcaption}
\usepackage{pgfplots}
\usepackage{arydshln}

\begin{document}

\title[Leveraging Visibility Graphs for Enhanced Arrhythmia Classification]{Leveraging Visibility Graphs for Enhanced Arrhythmia Classification with Graph Convolutional Networks}


\author[1]{\fnm{Rafael} \sur{F. Oliveira}}\email{rafael.fo@aluno.ufop.edu.br}
\author[1]{\fnm{Gladston} \sur{J. P. Moreira}}\email{gladston@ufop.edu.br}

\author*[1]{\fnm{Vander} \sur{L. S. Freitas}}\email{vander.freitas@ufop.edu.br}
\equalcont{These authors contributed equally to this work.}
\author[1]{\fnm{Eduardo} \sur{J. S. Luz}}\email{eduluz@ufop.edu.br}
\equalcont{These authors contributed equally to this work.}

\affil[1]{\orgdiv{Department of Computing}, \orgname{Federal University of Ouro Preto}, \orgaddress{\street{122, Diogo de Vasconcelos Street, Pilar}, \city{Ouro Preto}, \postcode{35402163}, \state{Minas Gerais}, \country{Brazil}}}


\abstract{Arrhythmias, detectable through electrocardiograms (ECGs), pose significant health risks, underscoring the need for accurate and efficient automated detection techniques. While recent advancements in graph-based methods have demonstrated potential to enhance arrhythmia classification, the challenge lies in effectively representing ECG signals as graphs. This study investigates the use of Visibility Graph (VG) and Vector Visibility Graph (VVG) representations combined with Graph Convolutional Networks (GCNs) for arrhythmia classification under the ANSI/AAMI standard, ensuring reproducibility and fair comparison with other techniques. Through extensive experiments on the MIT-BIH dataset, we evaluate various GCN architectures and preprocessing parameters. Our findings demonstrate that VG and VVG mappings enable GCNs to classify arrhythmias directly from raw ECG signals, without the need for preprocessing or noise removal. Notably, VG offers superior computational efficiency, while VVG delivers enhanced classification performance by leveraging additional lead features. The proposed approach outperforms baseline methods in several metrics, although challenges persist in classifying the supraventricular ectopic beat (S) class, particularly under the inter-patient paradigm.}




\maketitle

\section{Introduction}
\label{sec-intro}

Early detection of heart diseases is essential to enable preventive measures. The electrocardiogram (ECG)~\citep{cohen2019biomedical} is the primary diagnostic tool for heart conditions. Arrhythmias are heart issues detectable via ECG that impair the heart's ability to pump sufficient blood, potentially affecting the brain and other organs. Arrhythmias can be life-threatening, necessitating continuous cardiac activity monitoring for appropriate medical response \citep{pandey2019automatic, cheng2017life}.

The intricate and challenging task of manually identifying and classifying arrhythmias highlights the essential need for further research into automated solutions. Techniques involving Machine Learning, such as Artificial Neural Networks (ANNs) and Deep Neural Networks (DNNs), have been increasingly adopted \citep{luz2016ecg, zaoralek2018patient, hannun2019cardiologist}. However, achieving stringent and rigorous outcomes remains challenging, particularly within the inter-patient paradigm, where training and testing sets include distinct patient data \citep{aami:2008, de2004automatic, luz2016ecg}. In contrast, the intra-patient paradigm may encounter overestimated outcomes when training and testing sets consist of data from the same patient. \citet{de2004automatic} have articulated the importance of standard protocols for the sake of comparability among published studies, highlighting that those not adhering to such benchmarks might not fully realize their practical potential and could present results with a degree of bias \citep{luz2011choice, luz2016ecg}.

Numerous neural network-based methodologies proposed in the literature aim to address the automated classification of arrhythmias~\citep{mousavi2019inter, essa2021ensemble}. Studies such as those conducted by \citet{essa2021ensemble} and \citet{mousavi2019inter} have shown significant success in adhering to the AAMI \citep{aami:2008} standards and the inter-patient paradigm. Both used Convolutional Neural Networks (CNNs) in their machine learning models, albeit with distinct approaches. \citet{essa2021ensemble} proposed a hybrid architecture that combined CNN-LSTM and Long short-term memory (LSTM) networks using heartbeat segments, whereas \citet{mousavi2019inter} introduced a Recurrent Neural Network (RNN) structure consisting of an encoder-decoder with a CNN, applied to sequences of heartbeats. \citet{garcia2017inter} even introduced an innovative approach utilizing a temporal vectorcardiogram with feature selection through complex networks paired with an SVM classifier. When tested under the inter-patient paradigm, this method achieved an overall accuracy of 92.4\%.

Despite these advancements, exploring and representing ECG signals as graphs and applying Graph Neural Networks (GNNs) to this representation remains an underexplored area. This gap presents an opportunity for our proposal: leverage the Visibility Graph (VG) algorithm to transform ECG data into a graph structure, enabling GNNs. This approach combines the inherent advantages of graph representation with the robust feature extraction capabilities of GNNs, potentially setting a new benchmark in ECG-based arrhythmia classification.

Focusing on the proposed approach of employing a VG for ECG signal representation, one question that encapsulates the core objective of the study can be formulated:

\begin{itemize}
    \item How does the integration of the visibility graph (VG) approach to map ECG signals into graph structures and the incorporation of multi-lead ECG data in vector visibility graph (VVG) representations influence the classification performance and diagnostic accuracy of graph neural networks in identifying cardiac arrhythmias within the inter-patient evaluation protocol?    
\end{itemize}

Interestingly, our results show that simpler GCN architectures yielded better results than more complex ones, suggesting that simplicity in GCN structures can more effectively capture essential data characteristics and avoid unnecessary noise. Moreover, our findings demonstrate that modeling the signal as a graph, particularly considering both leads, is pertinent and promising for ECG analysis. The study further highlights the significant impact of the inter-patient and intra-patient paradigms on classification performance.

\section{Background}
\label{sec-back}

Network science \citep{barabasi2013network} has proven its value in extracting significant insights from various domains, including sequential data, enabling the characterization of nonlinear dynamic behavior in diverse contexts. Researchers have proposed several approaches for determining the spatial connectivity of data over time through complex networks. As highlighted by \citet{ren2019vector}, there are three primary methodologies prominent in the literature:

\begin{itemize}
    \item Mapping of sequential data or pseudoperiodic time series into complex networks, where network vertices represent each cycle of the time series. The connectivity between vertices relies on the temporal similarity or correlation between cycles \citep{zhang2006complex, sun2014characterizing};
    \item Recurrence networks that treat phase space vectors as vertices, with connectivity between vertices determined by the distance of the corresponding vectors \citep{donner2010recurrence, donges2011nonlinear};
    \item Direct definition of time series data as network vertices, where connectivity among vertices is determined based on the temporal sequence of data points such as a visibility criteria \citep{lacasa2008time, luque2009horizontal, gotoda2017characterization} or temporal neighborhood  \citep{Freitas_et_al_2019}.
\end{itemize}

Among the commonly used approaches in the literature, this article focuses on mapping ECG signals into graphs using two methods: the VG method proposed by \citet{lacasa2008time} for univariate time series (ECG signals with one lead) and the VVG method proposed by \citet{ren2019vector} for multivariate time series (ECG signals with two leads). 

\subsection{Visibility Graph (VG)}
\label{sub-visibility_graph}

We conceptualize the samples from an ECG lead curve over time as a one-dimensional time series, which were transformed into a graph using the Visibility Graph (VG), as depicted in Figure~\ref{fig:visibility_graph}. In this representation, each point in the series becomes a node in the graph, and two nodes are connected if they satisfy a visibility criterion, as follows: given two points, $(t_a, y_a)$ and $(t_b, y_b)$, represented by vertices $a$ and $b$, where $t_a$ and $t_b$ represent time and $y_a$ and $y_b$ their associated values, a connection is established between them if
\begin{equation}
    \label{eq:criterio_visibilidade_vg}
    y_c < y_b + \Bigl( y_a - y_b \Bigl) \dfrac{t_b - t_c}{t_b - t_a}
\end{equation} 

\noindent for any points $(t_c, y_c)$. In simpler terms, if each point in the series is a vertical bar in a bar chart, two bars $a$ and $b$ would connect if there is no bar $c$ between them whose height would prevent drawing a straight line between the peaks of $a$ and $b$.

\begin{figure*}[!ht]
  \centering
    \begin{tikzpicture}
        \begin{axis}[
            ybar,
            bar width=10pt,
            ymin=-0.2, ymax=0.7,
            xtick=data,
            xticklabels={0, 1, 2, 3, 4, 5, 6, 7, 8, 9, 10, 11, 12, 13, 14, 15, 16},
            xlabel={},
            ylabel={},
            width=15cm, 
            height=6cm,
            ]
            \addplot [ybar, fill=black] coordinates {(0,0.05) (1,-0.1) (2,-0.1) (3,-0.03) (4,0.25) (5,0.45) (6,0.6) (7,0.65) 
            (8,0.53) (9,0.27) (10,-0.1) (11,-0.15) (12,-0.15) (13,-0.1) (14,-0.05) (15,-0.03) (16,0.03)};
        \end{axis}
    \end{tikzpicture}        
    \label{fig:time_series}
\end{figure*}

\begin{figure*}[!ht]
    \centering
    \begin{tikzpicture}
        \begin{axis}[
            width=15cm, 
            height=6cm,
            xlabel={}, 
            ylabel={}, 
            xmin=-1.5, xmax=17.5, 
            ymin=-0.2, ymax=0.7, 
            xtick=data,
            xticklabels={0, 1, 2, 3, 4, 5, 6, 7, 8, 9, 10, 11, 12, 13, 14, 15, 16},,
        ]
        
        \addplot[
            only marks,
            mark=*,
            mark options={black},
        ] coordinates {
            (0, 0.05) (1, -0.1) (2, -0.1) (3, -0.03) (4, 0.25)
            (5, 0.45) (6, 0.6) (7, 0.65) (8, 0.53) (9, 0.27)
            (10, -0.1) (11, -0.15) (12, -0.15) (13, -0.1)
            (14, -0.05) (15, -0.03) (16, 0.03)
        };

        \addplot[
            thick,
            black 
        ] coordinates {
            (0, 0.05) (1, -0.1) (2, -0.1) (3, -0.03) (4, 0.25)
            (5, 0.45) (6, 0.6) (7, 0.65) (8, 0.53) (9, 0.27)
            (10, -0.1) (11, -0.15) (12, -0.15) (13, -0.1)
            (14, -0.05) (15, -0.03) (16, 0.03)
        };

        \draw[gray!70] (axis cs:0, 0.05) -- (axis cs:2, -0.1);
        \draw[gray!70] (axis cs:0, 0.05) -- (axis cs:3, -0.03);
        \draw[gray!70] (axis cs:0, 0.05) -- (axis cs:4, 0.25);
        \draw[gray!70] (axis cs:0, 0.05) -- (axis cs:5, 0.45);
        \draw[gray!70] (axis cs:0, 0.05) -- (axis cs:6, 0.6);
        \draw[gray!70] (axis cs:1, -0.1) -- (axis cs:3, -0.03);
        \draw[gray!70] (axis cs:1, -0.1) -- (axis cs:4, 0.25);
        \draw[gray!70] (axis cs:1, -0.1) -- (axis cs:5, 0.45);        
        \draw[gray!70] (axis cs:1, -0.1) -- (axis cs:6, 0.6);
        \draw[gray!70] (axis cs:2, -0.1) -- (axis cs:4, 0.25);
        \draw[gray!70] (axis cs:2, -0.1) -- (axis cs:5, 0.45);
        \draw[gray!70] (axis cs:7, 0.65) -- (axis cs:3, -0.03);
        \draw[gray!70] (axis cs:7, 0.65) -- (axis cs:4, 0.25);
        \draw[gray!70] (axis cs:7, 0.65) -- (axis cs:5, 0.45); 
        \draw[gray!70] (axis cs:7, 0.65) -- (axis cs:9, 0.27); 
        \draw[gray!70] (axis cs:7, 0.65) -- (axis cs:10, -0.1); 
        \draw[gray!70] (axis cs:7, 0.65) -- (axis cs:16, 0.03);
        \draw[gray!70] (axis cs:7, 0.65) -- (axis cs:14, -0.05);
        \draw[gray!70] (axis cs:7, 0.65) -- (axis cs:15, -0.03);
        \draw[gray!70] (axis cs:8, 0.53) -- (axis cs:11, -0.15);
        \draw[gray!70] (axis cs:8, 0.53) -- (axis cs:12, -0.15);
        \draw[gray!70] (axis cs:8, 0.53) -- (axis cs:13, -0.1);
        \draw[gray!70] (axis cs:8, 0.53) -- (axis cs:14, -0.05);
        \draw[gray!70] (axis cs:8, 0.53) -- (axis cs:15, -0.03);
        \draw[gray!70] (axis cs:8, 0.53) -- (axis cs:16, 0.03);
        \draw[gray!70] (axis cs:9, 0.27) -- (axis cs:11, -0.15);
        \draw[gray!70] (axis cs:9, 0.27) -- (axis cs:12, -0.15);
        \draw[gray!70] (axis cs:9, 0.27) -- (axis cs:13, -0.1);
        \draw[gray!70] (axis cs:9, 0.27) -- (axis cs:14, -0.05);
        \draw[gray!70] (axis cs:9, 0.27) -- (axis cs:15, -0.03);
        \draw[gray!70] (axis cs:9, 0.27) -- (axis cs:16, 0.03);
        \draw[gray!70] (axis cs:10, -0.1) -- (axis cs:13, -0.1);
        \draw[gray!70] (axis cs:10, -0.1) -- (axis cs:14, -0.05);
        \draw[gray!70] (axis cs:10, -0.1) -- (axis cs:15, -0.03);
        \draw[gray!70] (axis cs:10, -0.1) -- (axis cs:16, 0.03);
        \draw[gray!70] (axis cs:11, -0.15) -- (axis cs:13, -0.1);
        \draw[gray!70] (axis cs:11, -0.15) -- (axis cs:14, -0.05);
        \draw[gray!70] (axis cs:11, -0.15) -- (axis cs:16, 0.03);
        \end{axis}
    \end{tikzpicture}
    \caption{Example of the VG method application.}
    \label{fig:visibility_graph}
\end{figure*}

According to \citet{lacasa2008time}, the graph generated using the VG method consistently demonstrates certain intrinsic features. First, it is a connected graph wherein each vertex maintains visibility to its immediate neighbors, encompassing both left and right adjacencies. Second, the graph is undirected, a characteristic inherent to the algorithm's construction. Finally, the graph maintains invariance under data transformations of the series, with the visibility criterion remaining unaffected by horizontal and vertical axis rescaling and translations.

\subsection{Vector Visibility Graph (VVG)}
\label{sub-vector_visibility_graph}

Expanding on VG method for univariate time series, \citet{ren2019vector} proposed the Vector Visibility Graph (VVG) method for mapping multivariate time series into a directed complex network. This approach defines a multidimensional data vector as a node and establishes connectivity between nodes based on the visibility criterion applied to the corresponding data vectors.

Consider $X_t=\{x^i_t\}^m_{i=1}$, a multidimensional $m$-dimensional time series\footnote{Multidimensional time series representation for $m$ dimensions: $X_t=[\{x^1_{t}, x^1_{t+1}, x^1_{t+2},..., x^1_{t+n}\}, \{x^2_{t}, x^2_{t+1}, x^2_{t+2},..., \newline x^2_{t+n}\}, ..., \{x^m_{t}, x^m_{t+1}, x^m_{t+2},..., x^m_{t+n}\}]$}, where $n$ is the size of each dimension and ${\vec{X}_t}$ is a vector representing the space of the multivariate time series, given by $\vec{X}_t=[x^1_t, x^2_t, ..., x^m_t]$. For any two vectors $\vec{X}_a$ and $\vec{X}_b$ in the vector space ${\vec{X}_t}$, the projection of $\vec{X}_a$ onto $\vec{X}_b$ can be defined as follows:
\begin{equation}
\label{eq:projecao}
\lvert\lvert\vec{X}^a_b\rvert\rvert = \dfrac{\sum_{i=1}^{m} x^i_ax^i_b}{\sqrt{\sum_{i=1}^{m} x^i_ax^i_a}}.
\end{equation}

\begin{figure*}[!ht]
    \centering
    \begin{subfigure}{.45\textwidth}
      \centering
      \begin{subfigure}{\textwidth}
          \centering
          \includegraphics[width=1.2\textwidth]{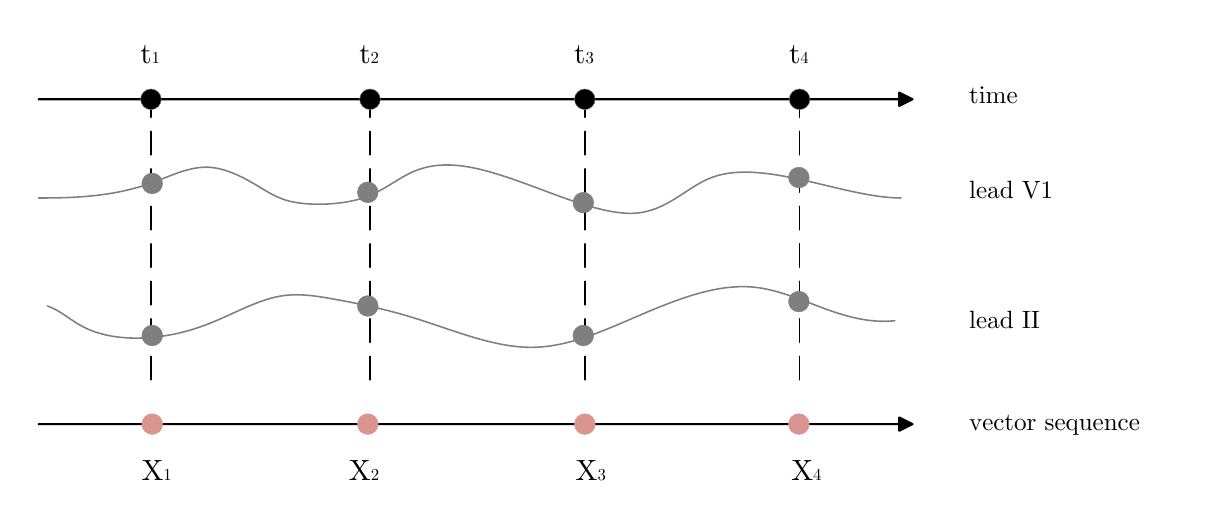}
          \caption{}
          \label{fig:time_series_vvg}
      \end{subfigure}
      \vfill
      \begin{subfigure}{\textwidth}
          \centering
          \includegraphics[width=1.2\textwidth]{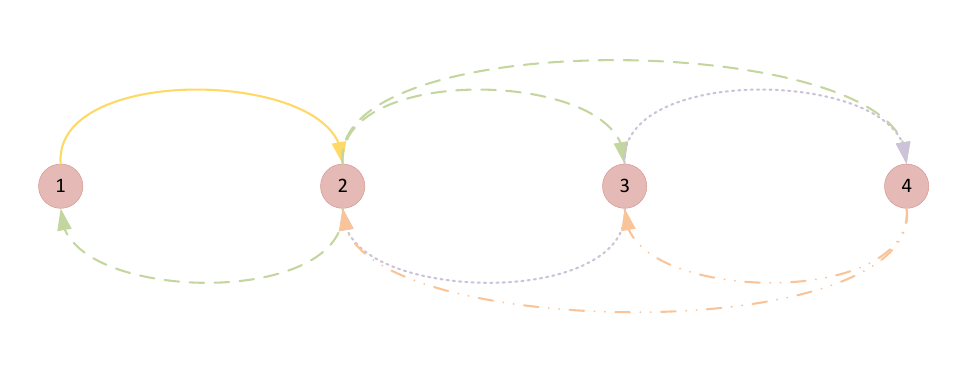}
          \caption{}
          \label{fig:projection_vvg}
      \end{subfigure}      
    \end{subfigure}
    \hfill \hfill
    \begin{subfigure}[b]{.5\textwidth}
      \centering
      \includegraphics[width=0.85\linewidth]{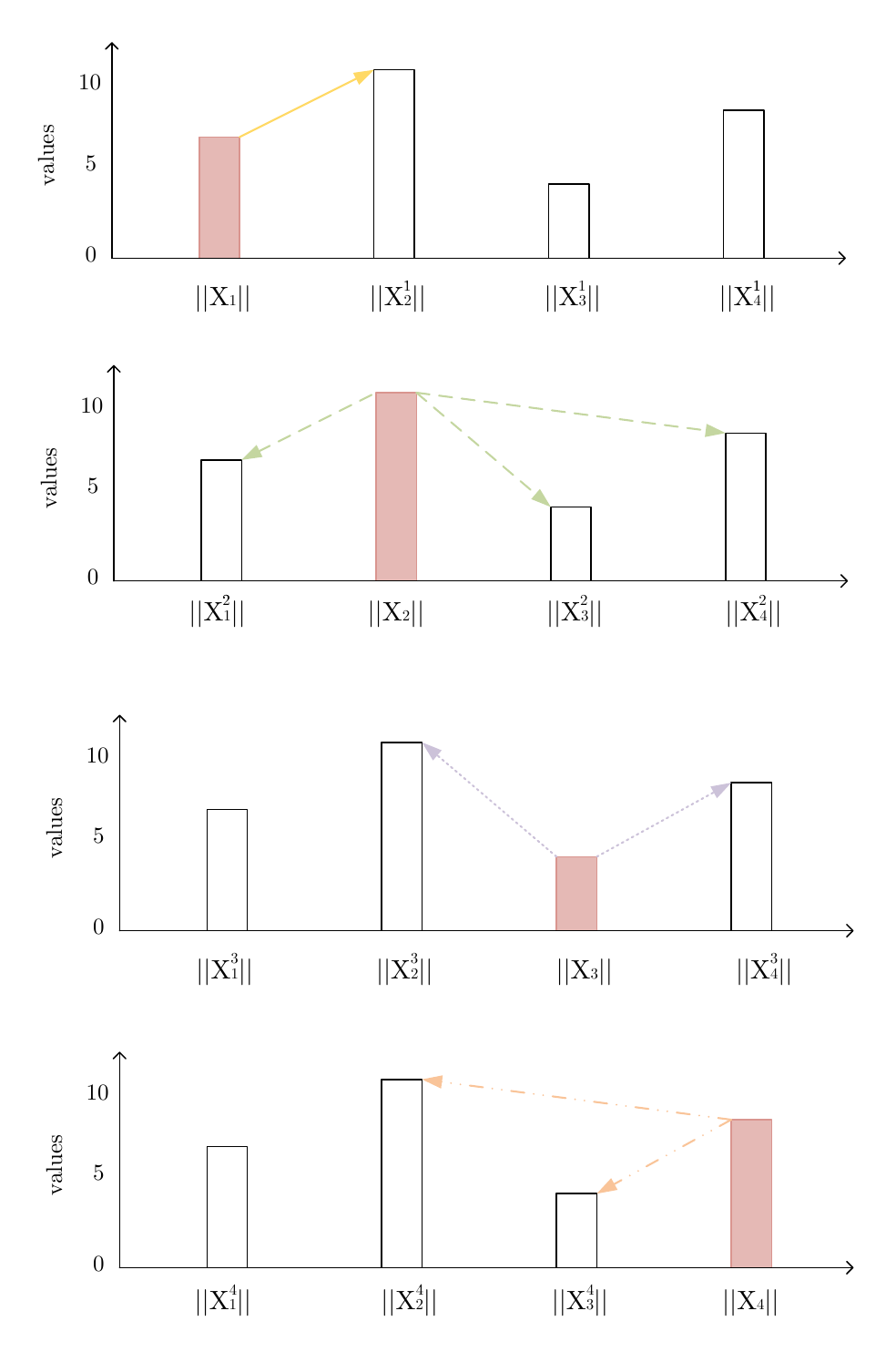}
      \caption{}
      \label{fig:vvg_graph}
    \end{subfigure}    
    \caption{Example of the VVG method application. (a) Multivariate time series represented as two ECG signals from different leads. (b) Graph generated by the VVG. (c) Visibility among the vectors. Source: Adapted from \citet{ren2019vector}.} 
    \label{fig:vvg_visibility}
\end{figure*}

Assuming each vector in the sequence represents a node in the network, we define the visibility criterion between vectors as follows:
\begin{equation}
\label{eq:criterio_visibilidade_vvg}
\lvert\lvert\vec{X}^a_c\rvert\rvert < \lvert\lvert\vec{X}^a_b\rvert\rvert + \Bigl( \lvert\lvert\vec{X}_a\rvert\rvert - \lvert\lvert\vec{X}^a_b\rvert\rvert \Bigl) \dfrac{t_b-t_c}{t_b-t_a},
\end{equation}

\noindent where $t_a < t_c < t_b$, $\lvert\lvert\vec{X}^a_b\rvert\rvert$ represents the projection of $\vec{X}_a$ onto $\vec{X}_b$, and $\lvert\lvert\vec{X}^a_c\rvert\rvert$ is the projection of $\vec{X}_a$ onto $\vec{X}_c$. We establish a connection from the vertex represented by $\vec{X}_a$ to the vertex represented by $\vec{X}_b$ in the resulting complex network if the criterion \eqref{eq:criterio_visibilidade_vvg} is fulfilled.  When $m=1$, the VVG method becomes equivalent to VG. Figure~\ref{fig:vvg_visibility} demonstrates the application of VVG. In this illustrated example, multivariate time series data, such as dual-lead ECG signals, are represented by a graph structure in which each series point corresponds to a network node. The VVG criteria determine the visibility between these points and illustrate how connections form within the graph. This representation highlights the complex relationships within the multivariate data and underscores the potential of the VVG method in capturing complex patterns in biomedical signals like ECG.

\subsection{Rationale for using VG and VVG}
\label{sub-rationale}

The rationale behind employing the VG methodology for analyzing ECG signals in our application is motivated by the nature of the ECG waveforms. ECG signals are characterized by prominent and well-defined waves, notably the P wave, QRS complex, and T wave, reflecting specific electrical activities within the heart. These waves form patterns that can indicate various cardiac conditions, including arrhythmias.

The hypothesis driving this approach is that in a normal cardiac rhythm, especially observable in lead II of the ECG, the peaks of these waves (P, QRS, and T) will exhibit a specific pattern of interconnectivity. The VG method transforms these time-series data points into a network of vertices (representing the wave peaks) and edges (representing the visibility or direct line of sight between these peaks). One expects to observe a consistent pattern of connections between these peaks in a regular heartbeat.

Conversely, these patterns will likely deviate from the norm in an arrhythmic beat. Arrhythmias often manifest irregularities in the timing, sequence, and morphology of ECG waves. For instance, atrial fibrillation may be characterized by an irregular rhythm and the absence of P waves, whereas ventricular tachycardia may show aberrant QRS complexes. By applying the VG method, we hypothesize that these disruptions in the waveform will result in distinctly different graph structures compared with those derived from normal heartbeats. We investigate this difference in graph topology through GNNs. This approach could enhance the accuracy of detecting and classifying cardiac arrhythmias by leveraging the complex interplay of ECG wave characteristics, which are often challenging to discern through traditional time-series analysis methods.

\subsection{Graph Neural Networks}
\label{sub-gnn}

The emergence of GNNs dates back to the 1990s \citep{zhou2020graph}, starting with recursive neural networks \citep{sperduti1997supervised} applied to directed acyclic graphs. Subsequently, Recurrent Neural Networks (RNNs) \citep{scarselli2008graph} and Feed-Forward Neural Networks (FFNNs) \citep{micheli2009neural} are introduced for cyclic graphs. With the advancement of deep learning, mainly through CNNs \citep{lecun1998gradient}, GNNs have also undergone significant enhancements.

\citet{wu2020comprehensive} categorizes GNNs into four main types: Recursive GNNs (RecGNN), Graph Autoencoders (GAE), Spatial-Temporal GNNs (STGNN), and Convolutional GNNs (ConvGNN) or Graph Convolutional Networks (GCN). 

ConvGNNs are the primary approaches for automatic arrhythmia classification in this work. They extend the concept of convolutional operations from the Euclidean domain, typically represented in grid formats, to graph-structured data. The core idea is to generate a representation for a node \( v \) by aggregating its attributes \( \mathbf{x}_v \) and the attributes of its neighboring vertices \( \mathbf{x}_u \), where \( u \in N(v) \), in which $N(v)$ is the set of neighbors.

ConvGNNs employ multiple convolutional layers to extract high-level representations of nodes. In this process, a single convolutional layer aggregates features from the first-order neighbors of a node, two layers aggregate features from two-hop neighborhoods, and so on. Consequently, the more convolutional layers used, the more extensive the data aggregated from a node's neighbors. Thus, ConvGNNs play a central role in developing complex GNN models because they effectively capture and integrate local and extended neighborhood features \citep{wu2020comprehensive}.

As detailed by \citet{wu2020comprehensive}, ConvGNNs are primarily distinguished into two methodologies. The first, spectral-based, defines graph convolutions by introducing filters in the domain of graph signal processing \citep{shuman2013emerging}, conceptualizing graph convolution as a means to remove noise from graph signals. The spatial-based one builds upon the propagation concept from RecGNNs and focuses on defining graph convolutions through disseminating features across neighborhoods. This latter approach has seen rapid development due to its efficiency, flexibility, and general applicability \citep{kipf2016semi}.

In this study, we chose the spatial-based methodology for ConvGNNs as the foundation for our proposed automatic ECG signal classification method. This decision is grounded in several key factors that align with the specific requirements and characteristics of ECG data analysis, such as (i) localized feature learning because ECG signals exhibit localized features, such as particular waveforms and intervals, that are crucial for identifying arrhythmias. In addition,  (ii) convolution operations are highly efficient. ECG-derived graphs can exhibit diverse and complex connectivity patterns, reflecting the intricate nature of cardiac electrical activity. The spatial approach allows for efficient processing of these graphs, adapting to the unique topology of each ECG-derived graph without the need for complex spectral transformations. (iii) The spatially-based ConvGNNs are robust to graph structure and size variations, making them suitable for ECG signals, which can vary significantly between individuals. This robustness ensures the model can generalize well across different patients, enhancing its utility in real-world clinical settings. (iv) Spatial-based ConvGNNs operate by directly aggregating features from neighboring nodes, potentially offering clinicians more intuitive interpretations.

GCNs take as input a graph represented as \( G(P, C) \), where \( P \) is the set of vertices and \( C \) the set of edges. Each vertex connects to itself to include its attributes in the aggregation and its neighbors' attributes. The attribute matrix \( X \in \mathbb{R}^{n \times d} \) contains attribute vectors for each vertex. The adjacency matrix \( A \) and the diagonal degree matrix \( D \), with \( D_{ii}=\sum_j A_{ij} \), are modified by adding the identity matrix to \( A \) for loops, resulting in \( \tilde{A} = A+ \lambda I_n \). A GCN with multiple layers captures features from a wider range of neighborhoods, following the propagation rule \citep{kipf2016semi}:
\begin{equation}
    H^{(l+1)} = \sigma(\tilde{D}^{-\frac{1}{2}}\tilde{A}\tilde{D}^{-\frac{1}{2}}H^{(l)}W^{(l)}),
\end{equation}

\noindent where \( \tilde{A} \) is the adjacency matrix with loops, \( \tilde{D} \) the degree matrix with loops, \( W^{(l)} \) the weight matrix, and \( \sigma \) the activation function, such as ReLU, with \( H^{(0)}=X \).

Because the number of neighbors of a vertex can vary from one to thousands, it is sometimes inefficient to consider the entire neighborhood. For this problem, a network called GraphSAGE \citep{hamilton2017inductive} samples a fixed number of neighbors for each vertex. The convolution operation is given by:
\begin{equation}
    h_v^{(l)} = \sigma(W^{(l)} \cdot f_l(h_v^{(l-1)},\{h_u^{(l-1)}, \forall u \in S_{n(v)}\})),
\end{equation}

\noindent where \( f \) is an aggregation function, \( \sigma \) the activation function, \( S_{n(v)} \) the sampled neighbor of vertex \( v \) and \( h_v^{(l)}  \in \mathbb{R}^{d} \) the attribute vector of vertex \( v \) of the \( l \)-th layer with \( h_v^{(0)}=X_v \). GraphSAGE aggregates features from local neighbors, enhancing the classification process with each iteration. The aggregated features is concatenated and normalized in each iteration, incrementally enriching the vertex representation.

GNNs analyze different levels of graph tasks, as outlined by \citet{zhou2020graph}. Node-level tasks include classification, regression, and clustering of vertices. Edge-level tasks involve classifying or predicting edges, whereas graph-level tasks encompass classification and regression for the entire graph. This study maps heartbeats into graphs, and the graph-level learning mechanism classifies each heartbeat by considering the graph as a whole.

\section{Related Works}
\label{sec-works}

This section summarizes studies that explored deep learning techniques for the problem of arrhythmia classification in ECG signals, as seen in Table~\ref{tab:comparativo_dnn}.
\begin{table*}[!th]
    \centering
    \caption{State-of-the-art DL works for arrhythmia classification.}
    \label{tab:comparativo_dnn}
    \resizebox{\textwidth}{!}{%
    \begin{tabular}{llcllcc}
        \toprule[1pt]
        \textbf{Reference} & \textbf{Dataset} & \textbf{\# Classes} & \textbf{Method} & \textbf{Performance (\%)} & \textbf{AAMI} &\textbf{Inter-patient}\\
        \toprule[1pt]
        \citet{cao2023ecg} & MIT-BIH & 4 & Pre-trained ResNet18 & $Acc^* \Rightarrow 90.8$, $Pr_N \Rightarrow 95.3$, $Re_N \Rightarrow 95.1$ & $\pm^*$ & \checkmark \\
        &&&& $Pr_S \Rightarrow 13.0$, $Re_S \Rightarrow 9.0$, $Pr_V \Rightarrow 68.2$, &&\\
        &&&& $Re_V \Rightarrow 88.4$, $Pr_F \Rightarrow 1.3$, $Re_F \Rightarrow 0.3$&&\\ 
        \midrule           
        \citet{gai2022ecg} & MIT-BIH & 5 & Pre-trained ImageNet & $Pr^* \Rightarrow 98.62$, $Re^* \Rightarrow 98.65$, & $\pm$ & \checkmark\\
        &&& Wigner-Villee distribution &$F1^* \Rightarrow 98.62$, $Acc \Rightarrow 98.65$ &&\\ 
        \midrule
        \citet{essa2021ensemble} & MIT-BIH & 4 & CNN-LSTM & $Acc \Rightarrow 95.81$, $Sp^* \Rightarrow 94.56$, $Se \Rightarrow 69.20$,  & \checkmark & \checkmark\\
        & && RR-HOS-LSTM & $F1 \Rightarrow 71.06$, $+P^* \Rightarrow 74.97$, $\kappa^* \Rightarrow 0.79$ &&\\ 
        \midrule
        \citet{hannun2019cardiologist} & Own dataset & 12 & DNN & $AUC^* \Rightarrow 97.0$, $Sp \Rightarrow 75.2$ & X & \checkmark\\
        &&&& $F1 \Rightarrow 83.7$ &&\\ 
        \midrule
        \citet{mousavi2019inter} & MIT-BIH & 4 & Sequence-to-sequence CNN & $Acc \Rightarrow 99.53$, $Se \Rightarrow 96.18$ & \checkmark & \checkmark\\
        &&&& $+P \Rightarrow 97.2$, $Sp \Rightarrow 98.58$ &&\\ \midrule
        \citet{garcia2017inter} & MIT-BIH & 3 & TVCG* + Complex Networks and SVM & $Acc \Rightarrow 92.4$ & \checkmark & \checkmark\\  
        \midrule        
        \citet{mathews2018novel} & MIT-BIH & 2 & Restricted Boltzmann Machines (RBM) & $Acc \Rightarrow 95.2$, $Se \Rightarrow 80.5$ & \checkmark & \checkmark \\
        &&&Deep Belief Network (DBN) & $+P \Rightarrow 47.97$, $FPR \Rightarrow 4.65$ &&\\
        \botrule
        \end{tabular}}
    {\footnotesize $^*Acc \Rightarrow$ Accuracy, $Se \Rightarrow$ Sensitivity, $Sp \Rightarrow$ Specificity, $+P \Rightarrow$ Positive Prediction, $FPR \Rightarrow$ False Positive Rate, $Pr \Rightarrow$ Precision, $Re \Rightarrow$ Recall, $F1 \Rightarrow$ F1-score, $AUC \Rightarrow$ Area Under the Curve, $\kappa \Rightarrow$ Kappa, $\pm \Rightarrow$ Partially met AAMI standard, $TVCG \Rightarrow$ Temporal Vectorcardiogram.}
\end{table*}

\citet{hannun2019cardiologist} introduces a landmark development in arrhythmia classification through ECG signals, utilizing a basic CNN architecture. This study stands out because it uses a large private dataset encompassing ECGs from 53,549 patients, demonstrating that deep learning techniques can surpass even cardiologists in arrhythmia detection and classification. The innovative approach of using single-channel ECG signals without extensive preprocessing highlights the effectiveness of deep learning despite data and signal limitations. However, the dataset's exclusivity limits the results' replicability, in contrast with the ANSI/AAMI \citep{aami:2008} norm recommending the use of public datasets, such as the widely recognized MIT-BIH, for more standardized and comparable evaluations.

Studies employing the MIT-BIH Arrhythmia Database for evaluation, such as in \citep{gai2022ecg, mathews2018novel, cao2023ecg, essa2021ensemble}, are distinguished for adhering to the ANSI/AAMI standard, ensuring reproducibility and fair comparisons with other techniques. Nevertheless, \citet{de2004automatic} identified a lack of standardization in using the MIT-BIH, particularly when not employing an inter-patient scheme for evaluation. Intra-patient schemes, such as cross-validation, may yield clinically unrealistic results. For example, \citet{gai2022ecg} does not strictly follow the De Chazal et al. protocol and uses a limited number of beats for testing, making comparisons with other methods challenging. This lack of standardization and non-adherence to the protocol underscores the need to reevaluate and reimplement other methods for a fair comparison.

State-of-the-art studies following the De Chazal et al. protocol and the ANSI/AAMI standards include \citep{mathews2018novel, cao2023ecg, essa2021ensemble, mousavi2019inter}. These studies are distinguished by using deep learning in a fair and robust approach, employing an inter-patient scheme in constructing training and test datasets. For instance, \citet{mousavi2019inter} study used a sequence-to-sequence approach, processing multiple heartbeats simultaneously, which is crucial for inferring cardiac rhythm from a chain of beats. The model combines CNN for feature extraction and RNN with LSTM units for encoding and decoding these sequences. The results demonstrate the proposed model's superior performance compared with other algorithms, achieving high accuracy, sensitivity, and positive predictive value for the analyzed arrhythmia categories in both intra-patient and inter-patient paradigms.

In \citet{mathews2018novel}, a novel approach is proposed for the classification of ECG signals using deep learning, explicitly employing Restricted Boltzmann Machines (RBM) and Deep Belief Networks (DBN) for detecting ventricular and supraventricular arrhythmias. This study utilizes the MIT-BIH database and conducts preprocessing of ECG signals through filtering artifacts such as baseline wander, power line interference, and high-frequency noise. The authors implemented two feature extraction techniques to produce feature sets: RR intervals, heartbeat intervals, and segmented morphology. The results demonstrate high accuracy in detecting ventricular ectopic beats (93.63\%) and supraventricular ectopic beats (95.57\%).

\citet{essa2021ensemble} proposes an automatic system for cardiac arrhythmia classification using deep learning models that combine CNN and LSTM to capture local features and temporal dynamics in ECG data. This model integrates classic features such as RR intervals and Higher Order Statistics (HOS) with the LSTM model to highlight abnormal heartbeat classes effectively. These models are trained on different data subsets to address class imbalance and then combined using a meta-classifier. Another model further verifies the outcome of the meta-classifier to reduce false positives. Experimental results obtained from the MIT-BIH Arrhythmia Database and following a ``subject-oriented (a.k.a. intra-patient)" independent patient evaluation scheme revealed that the proposed method achieves an overall accuracy of 95.81\%. The average F1 score and positive predictive value exceeded all other methods by over 3\% and 8\%, respectively.

The study in \citet{cao2023ecg} introduces a method for cardiac arrhythmia classification using deep learning with the ResNet-18 model. The preprocessing of the MIT-BIH data included filtering to remove noise and segmentation into heartbeats. The method employs a Short-Time Fourier Transform (STFT) to convert 1D ECG signals into 2D time-frequency spectrograms suitable for pre-trained CNN classifiers. This study uses oversampling and undersampling techniques to address class imbalance in the dataset. The results indicate an accuracy of 90.8\%, with high precision and recall for the normal class (N) and varied performance for other types of arrhythmias, such as supraventricular (S), ventricular (V), and fusion (F).

The present article explores deep learning techniques based on graphs, specifically GNNs, for which transforming ECG signals and heartbeats into graph representations is essential. Prior research has investigated the transformation of ECG signals into graphs, such as \citet{garcia2017inter}. They introduced an ECG representation based on a Temporal Vectorcardiogram (TVCG), coupled with a complex network for feature extraction and resource selection using a Particle Swarm Optimization (PSO) algorithm. This approach finely tunes an SVM classifier. The proposed method proved effective in the inter-patient paradigm, with results comparable to the state-of-the-art in the MIT-BIH database, achieving 53\% positive predictivity for the supraventricular ectopic beat class and 87.3\% sensitivity for the ventricular ectopic beat class. The TVCG is a set of points representing two derivations and time in this method. They build a network by considering these points as vertices and using the Euclidean distance between each pair of points as edges, forming a square matrix. They initially transform each beat into a regular network (where all vertices are connected) and construct the graph by dynamically removing edges based on a threshold chosen during training. These graphs extract features, including the Mean connectivity degree, Maximum connectivity degree, Joint degree entropy, Joint degree energy, and Mean joint degree. An SVM classifier is then trained, thus not exploring GNNs for the classification task.

The study in \citet{zhao2022ecgnn} presents an innovative methodology for recognizing abnormalities in 12-lead ECGs using a GNN. The methodology comprises two main modules: a feature extractor and a graph neural network module. The feature extractor transforms raw ECG signals into feature vectors and uses these vectors to initialize the graph nodes. The GCN module then performs convolution and pooling operations on graphs to generate new subgraphs and graph-level representations. For graph construction, they represent each ECG sample with an arbitrary graph of 12 vertices, where each vertex represents a lead. A lead-level attention mechanism was also applied to highlight leads with features most relevant to specific heart diseases. The graph's adjacency matrix describes the connection features between the leads, and the node feature matrix represents the attributes of each lead. The evaluation used the PTB-XL and ICBEB2018 datasets containing 12-lead ECG records with various annotations and diagnostic categories.

In \citet{DUONG2023120107}, the classification of ECG signals is addressed using an approach that combines edge detection and GNN. This methodology encompasses two main phases: data transformation and classification. Initially, the ECG is processed as a \(64 \times 64 \times 1\) image to construct graph-formatted data. This process includes applying the Sobel operator for edge detection on the curve images. Each pixel with a grayscale intensity value of 128 or higher is converted into a graph node, and the intensity of that pixel becomes an attribute of the node. Edges connect vertices that group neighboring pixels. A graph is constructed from a single image with nodes and edges derived from the image. The grayscale intensity values of the vertices are normalized for each graph, which involves subtracting the mean of all attributes under each graph from the original value of the preprocessed image, followed by division by the standard deviation. Various graph network architectures are considered for evaluation and conducted on two datasets, the MIT-BIH and PTB-XL. The authors compare their approach with that proposed in \citet{zhao2022ecgnn}, claiming superiority as a 100\% result is achieved in various graph network architectures. However, inter-patient evaluation is not considered in both approaches proposed in \citet{DUONG2023120107} and \citet{zhao2022ecgnn}, which may explain the significant results. In addition, they do not consider the ANSI/AAMI EC57 standard recommendations for grouping heartbeat classes.

The evaluation we propose in our study adheres to the ANSI/AAMI EC57 standard and follows the inter-patient evaluation protocol. We also envision conducting experiments without preprocessing the signals or techniques to address class imbalance, as these techniques can significantly influence the final results. By adopting this strategy, we aim to assess the raw and inherent potential of GNNs in discerning arrhythmia patterns directly from ECG data.

Furthermore, we extend this concept by exploring multivariate visibility graphs to incorporate multiple ECG leads. This advancement aligns with clinical practices, where various leads are analyzed for comprehensive cardiac assessment and enriches the data representation, potentially capturing a more complete picture of cardiac rhythm. We have developed an implementation of this multivariate visibility graph method~\footnote{\url{https://github.com/raffoliveira/ts2vvg}} that enhances computational feasibility without compromising the integrity of the ECG signal representation. Compliance with the ANSI/AAMI EC57 standard and the inter-patient evaluation protocol ensures that our findings are not only scientifically rigorous but also hold practical relevance and can be reliably compared with existing methodologies. 

\section{Experimental methodology}
\label{sec-metho}

\begin{figure*}[!ht]
    \centering
    \includegraphics[width=.8\linewidth]{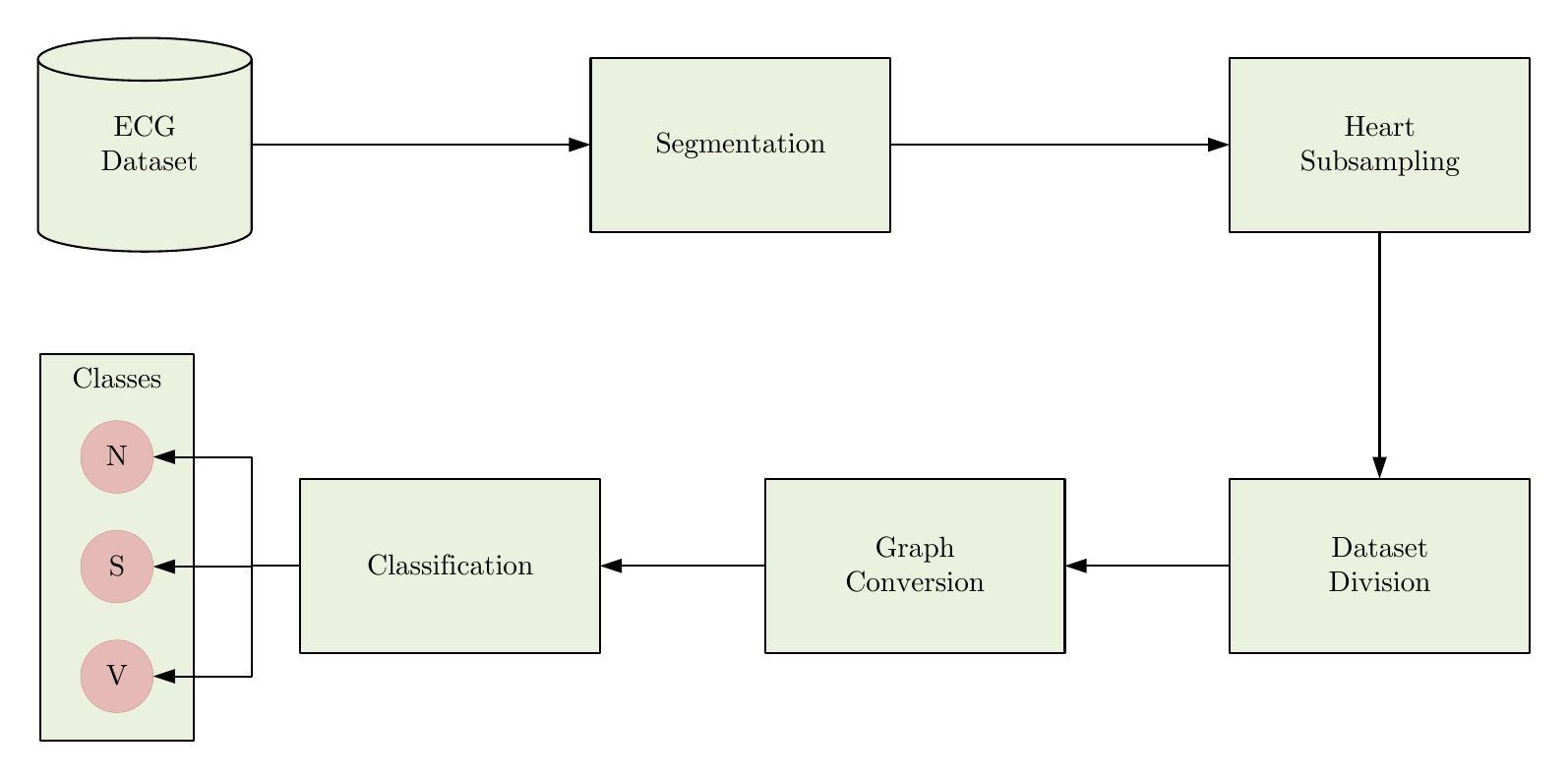}
    \caption{Flowchart of our proposed methodology.}
    \label{fig:methodology_diagram}
\end{figure*}

Here, we present the proposed methodology, summarized in Figure~\ref{fig:methodology_diagram}. We use the ECG signals in their original, unprocessed form without applying any noise-reducing filters.

\subsection{Dataset and ANSI/AAMI EC57}
\label{sec:database_aami}

Evaluation of automatic arrhythmia classification methods in the literature necessitates a dataset with heartbeats grouped into patient records. The Association for the Advancement of Medical Instrumentation (AAMI) developed a standard, outlined in ANSI/AAMI EC57:1998/(R)2008 \citep{aami:2008}, to standardize the evaluation of the methods, ensuring reproducibility and comparability. This standard recommends using one of the following five datasets:

\begin{itemize}
\item MIT-BIH: The Massachusetts Institute of Technology - Beth Israel Hospital Arrhythmia Database (48 records of 30 minutes each).
\item  CU: The Creighton University Sustained Ventricular Arrhythmia Database (35 records of 8 minutes each).
\item  AHA: The American Heart Association Database for Evaluation of Ventricular Arrhythmia Detectors (80 records of 35 minutes each).
\item  ESC: The European Society of Cardiology ST-T Database (90 records of two hours each).
\item  NST: The Noise Stress Test Database (12 records of 30 minutes each, plus three records with excessive noise).
\end{itemize}

The MIT-BIH dataset\footnote{https://physionet.org/content/mitdb/1.0.0/} is most representative in terms of arrhythmia types and is widely used in literature \citep{moody1990bih, moody2001impact}. This dataset, also used in this work, comprises 48 ECG signal records of 30 minutes from 47 patients sampled at 360 Hz. Each signal contains two leads.

AAMI's standard specifies annotation guidelines for each heartbeat in datasets, recommending excluding pacemaker data, segments with ventricular flutter or fibrillation (VF), and artificial data. We removed data from four patients and grouped the remaining into five main classes: Normal (N), Supraventricular ectopic beat (S), Ventricular ectopic beat (V), Fusion beat (F), and Unknown beat (Q).

The AAMI EC57 standard does not specify which data (patient heartbeats) should be used for training and testing classification models. Data division can follow either an intra-patient or inter-patient paradigm \citep{sraitih2021overview}. Intra-patient uses ECG signal data from the same patient for training and testing, whereas inter-patient involves data from different patients without overlap. Using data from the same patient in training and testing often leads to overestimated evaluations \citep{luz2011choice, luz2016ecg}.

To align tests with real-world scenarios, \citet{de2004automatic} proposed dividing the MIT-BIH dataset into two sets, DS1 and DS2, to ensure no overlap. Table \ref{tab:inter_patient} shows the distribution of patient records between the two sets.
\begin{table}[!ht]
    \centering
    \caption{Distribution of MIT-BIH patient records in two sets.}
    \label{tab:inter_patient}
    \begin{tabular}{cc}
    \toprule
        \textbf{ DS1} & \textbf{DS2}\\
        \midrule
        101, 106, 108, 109, 112,  & 100, 103, 105, 111, 113, \\
        114, 115, 116, 118, 119,  & 117, 121, 123, 200, 202, \\
        122, 124, 201, 203, 205,  & 210, 212, 213, 214, 219, \\ 
        207, 208, 209, 215, 220,  & 221, 222, 228, 231, 232, \\
        223, 230 & 233, 234 \\
        \bottomrule
    \end{tabular}
\end{table}

Using the inter-patient paradigm and the division proposed by De Chazal et al., Table \ref{tab:beat_mti_bih} summarizes the number of heartbeats per class and their percentage in the MIT-BIH dataset. The Table \ref{tab:beat_mti_bih} also shows the heartbeat counts after segmentation into training (DS1) and testing (DS2) sets, with percentages of 50,65\% and 49,35\%, respectively. Classes F and Q, with less than 1\% presence, are excluded from experiments because of their low occurrence.
\begin{table*}[!ht]
    \centering
    \caption{Description of the number of beats in the training and test sets.}
    \label{tab:beat_mti_bih}
    \begin{tabular}{lcccc}
        \toprule
        \textbf{Beats} & \textbf{\% of total} & \textbf{\# Training (DS1)} &
        \textbf{\# Testing (DS2)} &\textbf{\# Total} \\
        \midrule
        \textbf{N} & 89.47\% & 45844 & 44238 & 90082 \\
        \textbf{S} & 2.76\% & 944 & 1837 & 2781\\
        \textbf{V} & 6.96\% & 3788 & 3220 & 7008\\
        \textbf{F} & 0.80\% & 414 & 388 & 802\\
        \textbf{Q} & 0.01\% & 8 & 7 & 15\\
        \midrule
        \textbf{Total} & & \textbf{50998} & \textbf{49690}  & \textbf{100688}\\
        \cmidrule{3-5}
        \textbf{Percentage} & 100\% & 50.65\% & 49.35\% &\\
        \bottomrule
    \end{tabular}
\end{table*}

\begin{table}[!ht]
    \centering
    \caption{Description of subsampling strategy.}
    \label{tab:sampling}
    \begin{tabular}{cccccc}
        \toprule
        \multirow{3}{*}{\textbf{Beats}} & \multicolumn{2}{c}{\textbf{Training (DS1)}} && \multicolumn{2}{c}{\textbf{Testing (DS2)}}\\
        \cmidrule{2-6}
        & Before & After && Before & After\\
        \toprule
        N & 45844 & 4584 && 44238 & 4423\\
        S & 944 & 944 && 1837 & 1837\\
        V & 3788 & 3788 && 3220 & 3220\\ 
        \midrule
        \textbf{Total} & \textbf{50576} & \textbf{7732} && \textbf{49295} & \textbf{8057}\\
        \bottomrule
    \end{tabular}
\end{table}

\subsection{Segmentation}
\label{sec-segm}

ECG signal segmentation is carried out in sample windows using the R-wave location metadata provided with the MIT-BIH dataset. This involved capturing $y$ points before the R-wave and $y$ points after it, resulting in heartbeats comprising $Y$ points, where $Y = y_{\text{before}}+y_{\text{after}}$. We then label each heartbeat according to the classes annotated in the dataset and conduct experiments to determine the optimal value of $Y$ for this phase. In some experiments, we normalize ECG signals to a range between 0 and 1.

\subsection{Heartbeat Subsampling}

The preponderance of class N heartbeats, as illustrated in Table~\ref{tab:beat_mti_bih}, necessitates a substantial computational effort due to the numerous graphs required. A subsampling strategy is implemented for class N within the training (DS1) and testing (DS2) datasets to facilitate a broader range of experiments. A sampling rate of 10\% is adopted, selectively choosing the final heartbeat in every sequence of ten to enhance diversity. Table~\ref{tab:sampling} provides a comparative overview of the dataset pre- and post-balance, including the removal of the less prevalent classes F and Q.

\subsection{Dataset Division}

We divide the heartbeats segmented in the prior stage into a training set (DS1) and a testing set (DS2), as detailed in Table~\ref{tab:beat_mti_bih}. This division adheres to the inter-patient paradigm, aiming to mirror a more realistic scenario \citep{de2004automatic}. Furthermore, following the subsampling of the predominant class N, as shown in Table~\ref{tab:sampling}, it is observed that both the total number of heartbeats and the number in class S are higher in DS2 than in DS1. Consequently, we design two additional experiments to examine GCN model performance when reversing the roles of DS1 and DS2 and when dividing the data according to the intra-patient paradigm.

\subsection{Conversion of Heartbeats into Graphs}

\begin{figure*}[!ht]
   \centering
   \includegraphics[width=\linewidth]{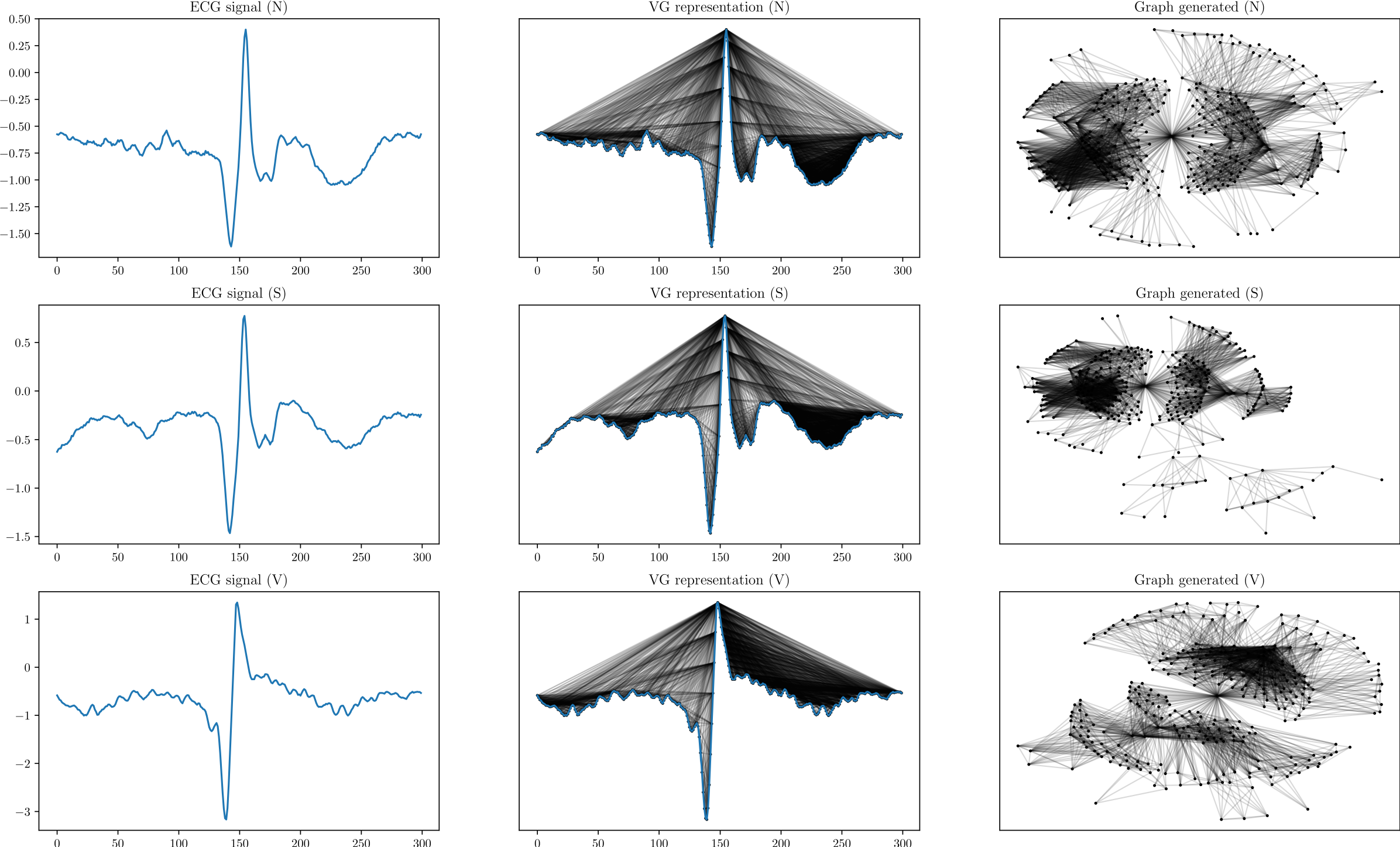}
   \caption{Example of mapping heartbeats from one lead (patient 118) corresponding to classes N, S, and V using the VG method. } 
   \label{fig:beats_graphs}
\end{figure*}

In this phase, we transform heartbeats into graphs using the VG and VVG methods. Figure~\ref{fig:beats_graphs} illustrates the conversion of a heartbeat from the MIT-BIH training set (DS1) corresponding to classes N, S, and V into a graph via the VG method for one ECG lead. This process converts a heartbeat \( j \) into a graph G.

This study employs a graph-level approach, where classification considers the graph as a whole, similar to the approach by \citet{kojima2020kgcn}, which used GCN to predict protein structures. The use of GNNs enables the incorporation of additional features into the nodes or edges of the graph. Experiments are designed to evaluate different sets of features' added to the nodes and their impact on the performance of GCNs in classifying arrhythmias from ECG signals.

\subsection{Classification}

We evaluate five GCN architectures for heartbeat classification alongside CNNs for comparative analysis. While CNNs are not the focus of this study, they provide a benchmark for comparing GCN performance.

In GCN architectures, the input layer takes the attribute vector size ($d$) of each node ($\mathbf{x}_v \in \mathbb{R}^d$), i.e., the amount of features aggregated at each node. The output layer features three neurons corresponding to the three target classes (N, S, V). For graph-level classification, a \textit{readout} layer \eqref{eq:mean_readout} aggregates attributes from all vertices in the last iteration of processing, both during training and testing:

\begin{equation}
    \label{eq:mean_readout}
    h_G = \text{readout}(\{h_v^{(l)} | v \in G\}),
\end{equation}

\noindent
where $h_v^{(l)}$ is the attribute vector representation of vertex $v$ at the $l$-th iteration/layer with $h_v^{(0)}=X$. The \textit{readout} function can be a simple invariant operation like sum or mean or a more sophisticated graph-level aggregation function. Here, the mean operation obtains a high-level representation of the entire graph ($h_G$) \citep{xu2018powerful}.

Following the methodology description, experiments are designed to investigate the primary question of this work: can graph representations of ECG signals using VG and VVG methods enhance arrhythmia classification performance using GNNs? The designed experiments are as follows:

\begin{itemize}
    \item \textbf{Experiment 1:} explore new GCN architectures using the VG method alongside various CNN architectures.
    \item \textbf{Experiment 2:} investigate different window sizes for ECG signal segmentation in the best GCN architectures using the VG method and the optimal CNN architecture.
    \item \textbf{Experiment 3:} examine features aggregation at graph vertices in the best GCN architectures using VG and VVG methods.
    \item \textbf{Experiment 4:} investigate the swap of datasets DS1 and DS2 in the best GCN architectures using VG and VVG methods and the optimal CNN architecture.
    \item \textbf{Experiment 5:} explore the intra-patient paradigm in the best GCN architectures using VG and VVG methods and the optimal CNN architecture.
    \item \textbf{Experiment 6:} compare the proposed methodology with \citet{garcia2017inter} work.
\end{itemize}

\section{Results and Discussion}
\label{sec:results}

This section presents the experiments' findings to address the research questions. The source code is reproducible at github repository \footnote{\url{https://github.com/raffoliveira/VG_for_arrhythmia_classification_with_GCN}}.

\subsection{Experiment 1: New Architectures}
\label{subsec:experiment_1}

This stage involves experimenting with various GCN and CNN architectures, the latter serving as a comparative baseline for arrhythmia classification performance. The aim is to assess the performance of the proposed GCN architectures against CNNs, which are widely used in the literature. Variations in GCN architectures included the types of layers (traditional graph convolutional layers (GraphConv) and specialized layers (SAGEConv)) and the number of neurons in hidden layers. Variations in CNN architectures focus on the number of convolutional layers.

\begin{table*}[!ht]
    \centering
    \caption{GCNs Architectures.}
    \resizebox{\textwidth}{!}{%
    \begin{tabular}{llc|llc|llc|llc|llc}
        \toprule
        \multicolumn{3}{c|}{\textbf{GCN2}} &  \multicolumn{3}{c|}{\textbf{GCN7}} & \multicolumn{3}{c|}{\textbf{GCN60}} & \multicolumn{3}{c|}{\textbf{GCN120}} & \multicolumn{3}{c}{\textbf{GCN240}}\\
        \toprule
        \textbf{\#} & \textbf{Layer} & \textbf{Shape} & \textbf{\#} & \textbf{Layer} & \textbf{Shape} & \textbf{\#} & \textbf{Layer} & \textbf{Shape} & \textbf{\#} & \textbf{Layer} & \textbf{Shape} & \textbf{\#} & \textbf{Layer} & \textbf{Shape}\\
        \midrule 
        1 & GraphConv & $d^{*}\times$20 & 1 & SAGEConv  & $d\times$20 & 1 & SAGEConv & $d\times$60 & 1 & SAGEConv & $d\times$120 & 1 & SAGEConv & $d\times$240\\
        2 & GraphConv & 20$\times$3 & 2 & GraphConv & 50$\times$40 & 2 & SAGEConv & 60$\times$50 & 2 & SAGEConv & 120$\times$40 & 2 & SAGEConv & 240$\times$140\\
        3 & Readout & - & 3 & SAGEConv & 40$\times$30 & 3 & SAGEConv & 50$\times$35 & 3 & SAGEConv & 40$\times$20 & 3 & SAGEConv & 140$\times$40\\
        4 & Softmax & - & 4 & SAGEConv & 30$\times$20 & 4 & SAGEConv & 35$\times$3 & 4 & SAGEConv & 20$\times$3 & 4 & SAGEConv & 40$\times$3\\
        & & & 5 & SAGEConv & 20$\times$10 & 5 & Readout & - & 5 & Readout & - & 5 & Readout & - \\
        & & & 6 & GraphConv & 10$\times$5 & 6 & Softmax & - & 6 & Softmax & - & 6 & Softmax & - \\
        & & & 7 & GraphConv & 5$\times$3 &&& &&& &&&\\
        & & & 8 & Readout & - &&& &&& &&& \\
        & & & 9 & Softmax & - &&& &&& &&&\\       
        \bottomrule
    \end{tabular}}\label{tab:new_gcn}
    {\footnotesize *$d$ indicates each vertex's attribute/features vector size.}
\end{table*}

\begin{figure*}[!ht]
    \centering
    \begin{subfigure}[b]{0.45\textwidth}
        \includegraphics[width=\linewidth]{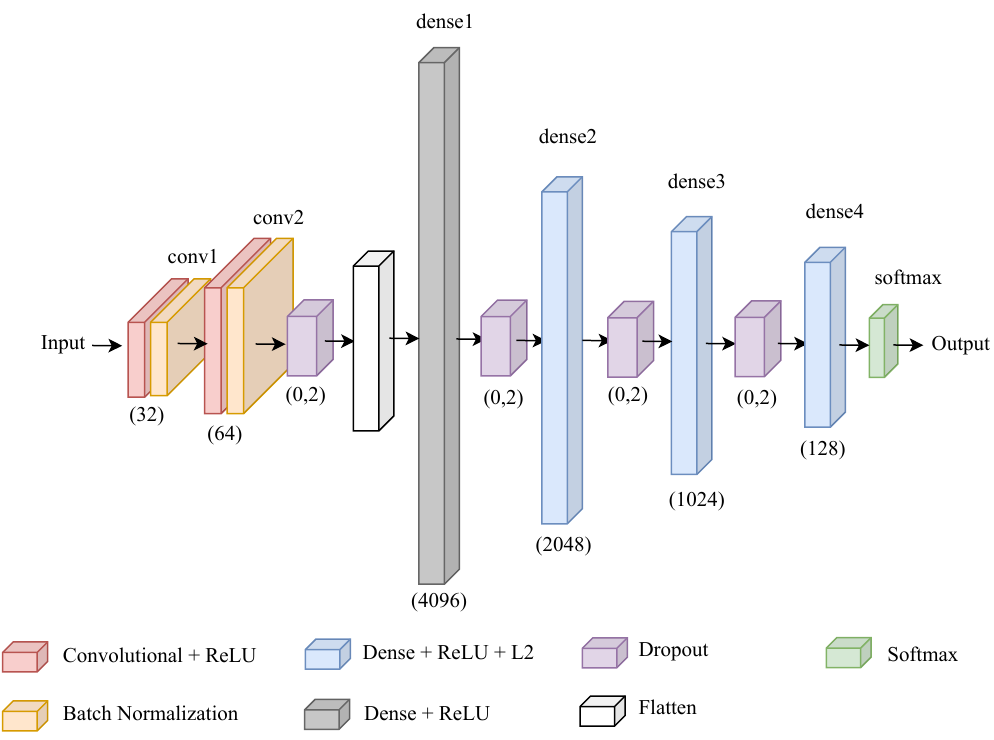}
        \caption{CNN-2Conv architecture.}
        \label{fig:cnn2conv}
    \end{subfigure}
    \begin{subfigure}[b]{0.45\textwidth}
        \includegraphics[width=\linewidth]{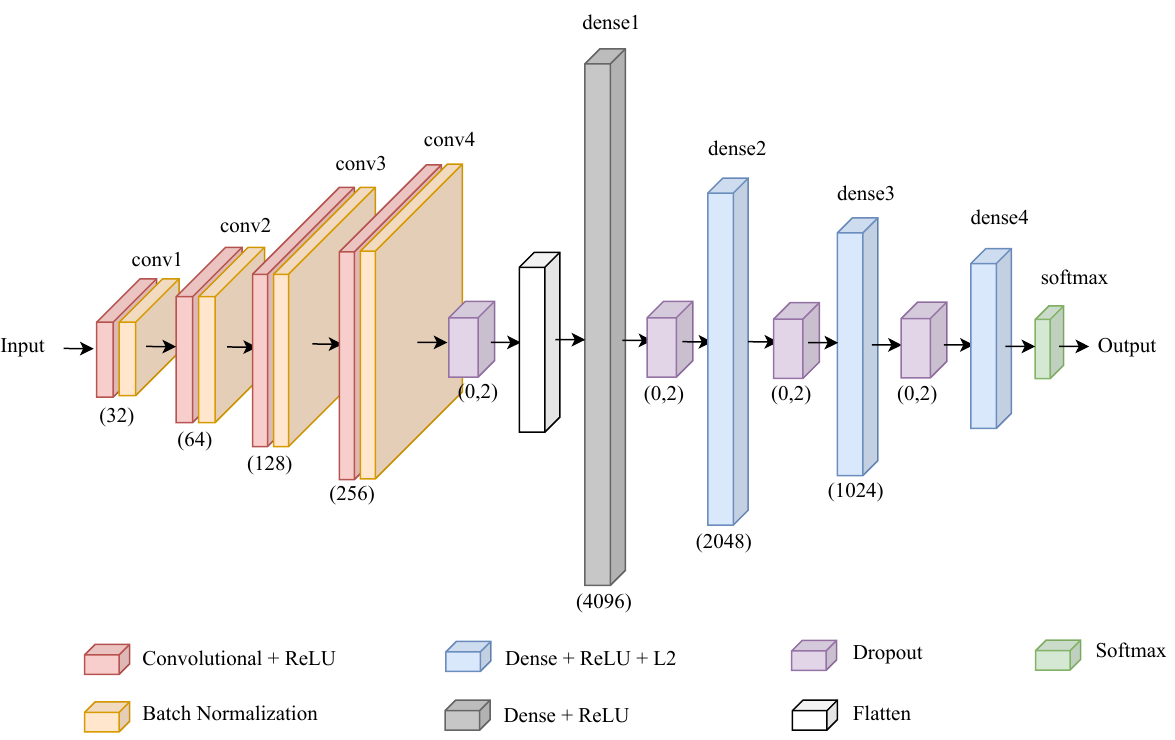}
        \caption{CNN-4Conv architecture.}
        \label{fig:cnn4conv}
    \end{subfigure}
    \begin{subfigure}[b]{.45\textwidth}
    \centering
        \includegraphics[width=\linewidth]{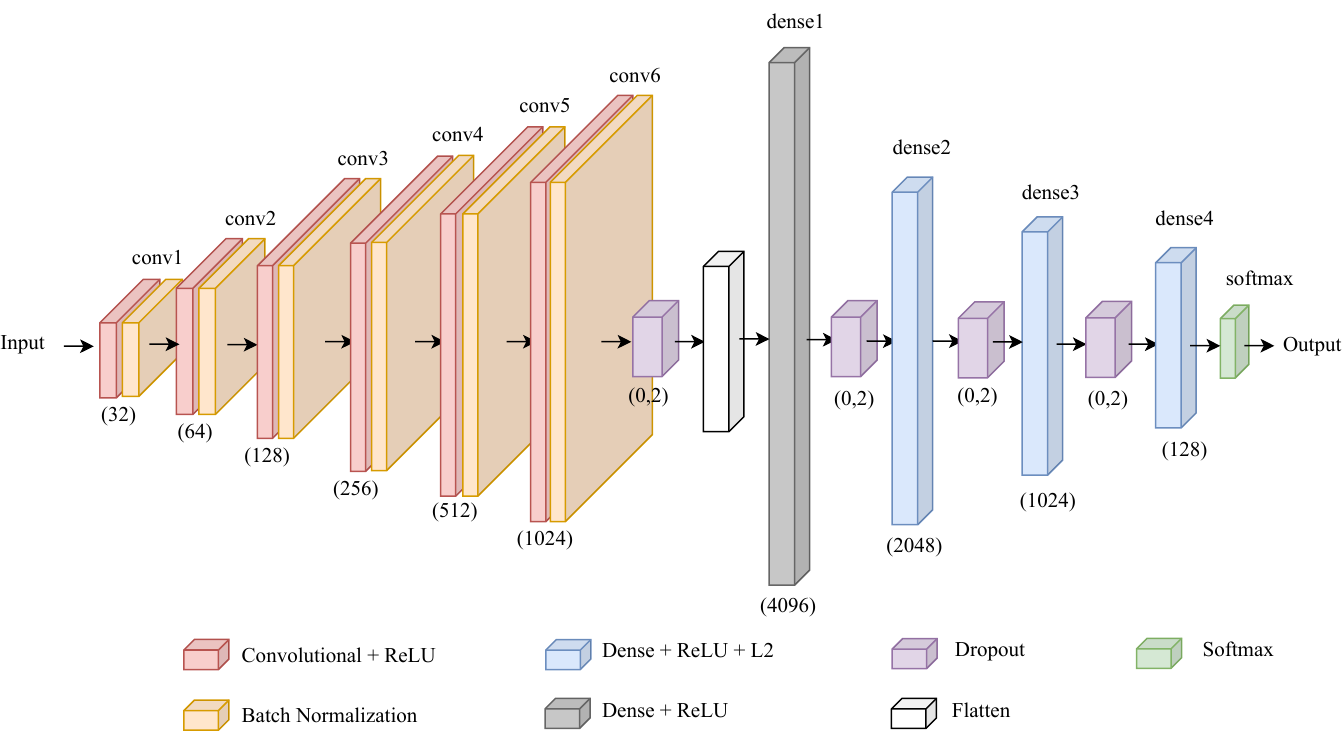}
        \caption{CNN-6Conv architecture.}
        \label{fig:cnn6conv}
    \end{subfigure}
    \caption{Proposed CNN architectures.}
    \label{fig:cnn_archi}
\end{figure*}

Table~\ref{tab:new_gcn} details the proposed GCN architectures for this experiment. Notably, a \textit{readout} layer is used to aggregate vertex attributes in the final iteration of processing to achieve a high-level representation of each entire graph, employing a mean operation. Figures \ref{fig:cnn2conv}, \ref{fig:cnn4conv}, and \ref{fig:cnn6conv} present the proposed CNN architectures, named CNN-2Conv, CNN-4Conv, and CNN-6Conv, respectively, based on the number of convolutional layers in each architecture. The CNN models employed 1D convolutional layers, with architectures based on an example of CNN use in electroencephalogram signal classification from the Keras library website\footnote{https://keras.io/examples/timeseries/eeg\_signal\_classification/}.

For the GCN experiments, each node is enriched with the following features: values from lead V1, values from lead II, and the relative timestamp of each point. This experiment used fewer aggregated data points to analyze different features groups in Experiment 3. Both sets of aggregated features are normalized between $[0, 1]$ using the min-max normalization technique\footnote{$X_{norm} = \frac{X - X_{min}}{X_{max} - X_{min}}$}. This technique is chosen to prevent the analyzed architectures from being influenced by extreme value data, as it preserves the order relationships among the data. No additional features is added to the CNN experiments, leaving the CNNs to ``learn" about the ECG signals independently. For both CNN training and converting ECG signals into graphs using the VG method for GNN training, lead II is used. A 10\% sampling of class N is performed, where only the last (tenth) heartbeat in a sequence of ten is chosen.

This experiment focused on parameter adjustments. Only the DS1 training set is used, divided into 80\% for training (DS1.1) and 20\% for validation (DS1.2). The inter-patient paradigm avoids data overlap between training and validation sets, with the distribution of MIT-BIH records as per Table~\ref{tab:DS1_inter_patient}. A fixed window of 280 points (100 points before and 180 points after the R peak) is used for segmenting ECG signals based on the state-of-the-art work of \citet{mousavi2019inter}.
\begin{table}[!ht]
    \centering
    \caption{Distribution of MIT-BIH records in the DS1 dataset into training (DS1.1) and validation (DS1.2) sets.}
    \label{tab:DS1_inter_patient}
    \begin{tabular}{cc}
    \toprule
        \textbf{Training (DS1.1)} & \textbf{Validation (DS1.2)} \\
        \midrule
        101, 106, 108, 112, 115, 116,  &  \\
        118, 119, 122, 124, 201, 203, & 109, 114, 207 e 223\\
        205, 208, 209, 215, 220 e 230 & \\
       \bottomrule
    \end{tabular}
\end{table}

In both GCN and CNN training, 150 epochs are used with the Adam optimizer at a learning rate 0.001. For CNN training, the Categorical Cross Entropy function \eqref{eq:loss_func} is the loss function. Thus, the CNN models continuously adjust their weights during the loss minimization, leading to better results. The mathematical expression that defines the loss function is:
\begin{equation}
    L(y_{i,\hat{y}_i}) = - \sum^L_{i=1} y_i*\log(\hat{y}_i),
    \label{eq:loss_func}
\end{equation}

\noindent
where $y_i$ is the actual label during training, $\hat{y}_i$ the predicted label, and $L$ the number of class labels.
\begin{table*}[!ht]
    \centering
    \caption{Summary of the performance of the GCN and CNN architectures.}
    \label{tab:results_GCN_CNN}
    \resizebox{\linewidth}{!}{
    \begin{tabular}{lccccccccccccccccccccc}
        \toprule
        \multirow{2}{*}{\textbf{Architectures}} & \multicolumn{4}{c}{\textbf{N}} && \multicolumn{4}{c}{\textbf{S}} && \multicolumn{4}{c}{\textbf{V}} && \multicolumn{4}{c}{\textbf{Weighted Average}} && \multirow{2}{*}{$\mathbf{Acc}^*$}\\
        \cmidrule{2-5} \cmidrule{7-10} \cmidrule{12-15} \cmidrule{17-20} 
         & $\mathbf{+P}^*$ & $\mathbf{Se}^*$ & $\mathbf{FPR}^*$ & $\mathbf{F_s}^*$ && $\mathbf{+P}$ & $\mathbf{Se}$ & $\mathbf{FPR}$ & $\mathbf{F_s}$ && $\mathbf{+P}$ & $\mathbf{Se}$ & $\mathbf{FPR}$ & $\mathbf{F_s}$ && $\mathbf{+P}$ & $\mathbf{Se}$ & $\mathbf{FPR}$ & $\mathbf{F_s}$ && \\
        \midrule \midrule
        \textbf{GCN2} & 47.0 & 62.0 & 58.26 & 53.0 && - & - & - & - && 40.0 & 36.0 & 43.02 & 38.0 && \textbf{39.0} & \textbf{44.0} & \textbf{45.18} & \textbf{41.0} && 44.0 \\
        \textbf{GCN7} & 67.0 & 46.0 & 18.31 & 55.0 && 33.0 & 1.0 & 0.26 & 2.0 && 55.0 & 86.0 & 55.25 & 67.0 && \textbf{58.0} & \textbf{59.0} & \textbf{32.5} & \textbf{54.0} && 59.0 \\
        \textbf{GCN60} & 35.0 & 43.0 & 64.29 & 39.0 && 1.0 & 1.0 & 6.52 & 1.0 && 32.0 & 28.0 & 46.28 & 30.0 && 30.0 & 32.0 & 50.05 & 31.0 && 32.0 \\
        \textbf{GCN120} & 37.0 & 47.0 & 64.64 & 42.0 && 1.0 & 1.0 & 5.02 & 1.0 && 35.0 & 31.0 & 44.75 & 33.0 && 33.0 & 35.0 & 49.37 & 33.0 && 35.0 \\
        \textbf{GCN240} & 36.0 & 43.0 & 62.24 & 39.0 && 4.0 & 3.0 & 8.18 & 3.0 && 32.0 & 29.0 & 46.99 & 31.0 && 31.0 & 32.0 & 49.62 & 31.0 && 32.0 \\
        \midrule\midrule
        \textbf{CNN-2Conv} & 82.0 & 50.0 & 8.89 & 62.0 && 43.0 & 16.0 & 2.64 & 23.0 && 60.0 & 94.0 & 48.83 & 73.0 && \textbf{70.0} & \textbf{65.0} & \textbf{25.69} & \textbf{63.0} && 65.0 \\
        \textbf{CNN-4Conv} & 77.0 & 49.0 & 11.82 & 60.0 && 67.0 & 14.0 & 0.84 & 23.0 && 57.0 & 90.0 & 52.6 & 70.0 && 67.0 & 63.0 & 28.47 & 60.0 && 63.0 \\
        \textbf{CNN-6Conv} & 81.0 & 43.0 & 8.16 & 56.0 && 79.0 & 16.0 & 0.52 & 26.0 && 55.0 & 93.0 & 58.61 & 70.0 && 68.0 & 62.0 & 29.41 & 59.0 && 62.0 \\
        \bottomrule
    \end{tabular}}
    {\footnotesize $^*Acc \Rightarrow$ Accuracy, $Se \Rightarrow$ Sensitivity, $+P \Rightarrow$ Positive Prediction, $FPR \Rightarrow$ False Positive Rate, $F_s \Rightarrow$ F1-score. \newline 
    \textbf{Note:} Values in bold indicate the best performance.}
\end{table*}

Given the results presented in Table~\ref{tab:results_GCN_CNN}, the two best GCN architectures and CNN architecture are chosen for further analysis in the following experiments. Table \ref{tab:results_GCN_CNN} summarizes the performance of architectures on the validation set (DS1.2), with GCN2 and GCN7 architectures showing the best results among GCNs and CNN-2Conv architecture yielding the best result among CNNs. The GCN2 and GCN7 architectures exhibited average $F_S$ of 41\% and 54\%, respectively. The CNN-2Conv architecture showed an average $F_S$ of 63\%. There is significant underperformance in class S for both architectures. As the architectures generally showed low performance, further experiments analyze other parameters and configurations to improve the current performance.

\subsection{Experiment 2: Investigating segmentation width}
\label{subsec:experiment_2}

This experiment investigates the influence of segmentation width, measured as the number of points per heartbeat, on the efficacy of automatic arrhythmia classification architectures. A critical aspect is that varying the segmentation width could lead to information loss from the ECG heartbeat data.

To this end, three distinct segmentation sizes — 230, 280, and 300 points per heartbeat — are scrutinized, each selected for their prevalence in the literature \citep{mousavi2019inter, mathunjwa2022ecg, ye2012heartbeat, shoughi2021practical, peimankar2021dens, zhao2005ecg}. The configurations for the dataset (DS1.1 and DS1.2) and the training protocols mirror those employed in Experiment 1.

As Table \ref{tab:results_segmentation} shows, the findings reveal a pronounced advantage at the 280-point segmentation width. Within the domain of GCN architectures, GCN7 stood out, registering an average $F_S$ of 54\%, while the CNN-2Conv model achieved an average $F_S$ of 63\%. Both architectures faced challenges in classifying class S, emphasizing a challenge in this class classification. Regarding segmentation width, GCN7 showed an upswing in performance by 20\% and 1.88\% for average $F_S$ compared to its counterparts at 230 and 300 points, respectively. Similarly, CNN-2Conv demonstrated enhancements of 23.53\% and 16.67\% in average $F_S$ relative to the 230 and 300-point windows, respectively. Consequently, the 280-point window has been identified as the optimal segmentation width for heartbeat analysis in this study stage.

\begin{table*}[!ht]
    \centering
    \caption{Summary of the GCN and CNN architectures' performance regarding the segmentation width.}
    \label{tab:results_segmentation}
    \resizebox{\linewidth}{!}{
    \begin{tabular}{lccccccccccccccccccccc}
        \toprule
        \multirow{2}{*}{\textbf{Architectures}} & \multicolumn{4}{c}{\textbf{N}} && \multicolumn{4}{c}{\textbf{S}} && \multicolumn{4}{c}{\textbf{V}} && \multicolumn{4}{c}{\textbf{Weighted Average}} && \multirow{2}{*}{$\mathbf{Acc}^*$}\\
        \cmidrule{2-5} \cmidrule{7-10} \cmidrule{12-15} \cmidrule{17-20} 
        & $\mathbf{+P}^*$ & $\mathbf{Se}^*$ & $\mathbf{FPR}^*$ & $\mathbf{F_s}^*$ && $\mathbf{+P}$ & $\mathbf{Se}$ & $\mathbf{FPR}$ & $\mathbf{F_s}$ && $\mathbf{+P}$ & $\mathbf{Se}$ & $\mathbf{FPR}$ & $\mathbf{F_s}$ && $\mathbf{+P}$ & $\mathbf{Se}$ & $\mathbf{FPR}$ & $\mathbf{F_s}$ && \\
        \midrule \midrule
        \multicolumn{22}{c}{\textbf{230 points}}\\
        \textbf{GCN2} & 43.0 & 53.0 & 57.53 & 47.0 && - & - & - & - && 37.0 & 37.0 & 50.25 & 37.0 && 36.0 & 40.0 & 48.01 & 38.0 && 40.0 \\
        \textbf{GCN7} & 49.0 & 45.0 & 33.63 & 47.0 && 5.0 & 3.0 & 14.07 & 4.0 && 48.0 & 59.0 & 48.62 & 53.0 && 44.0 & 46.0 & 38.04 & 45.0 && 46.0 \\
        \textbf{CNN-2Conv} & 59.0 & 43.0 & 24.9 & 50.0 && 40.0 & 15.0 & 2.7 & 21.0 && 50.0 & 72.0 & 56.07 & 59.0 && 53.0 & 52.0 & 36.1 & 51.0 && 52.0 \\
        \midrule
        \multicolumn{22}{c}{\textbf{280 points}}\\
        \textbf{GCN2} & 47.0 & 62.0 & 58.26 & 53.0 && - & - & - & - && 40.0 & 36.0 & 43.02 & 38.0 && \textbf{39.0} & \textbf{44.0} & \textbf{45.08} & \textbf{41.0} && 44.0 \\
        \textbf{GCN7} & 67.0 & 46.0 & 18.31 & 55.0 && 33.0 & 1.0 & 0.26 & 2.0 && 55.0 & 86.0 & 55.25 & 67.0 && \textbf{58.0} & \textbf{59.0} & \textbf{32.3} & \textbf{54.0} && 59.0 \\
        \textbf{CNN-2Conv} & 82.0 & 50.0 & 8.89 & 62.0 && 43.0 & 16.0 & 2.64 & 23.0 && 60.0 & 94.0 & 48.83 & 73.0 && \textbf{70.0} & \textbf{65.0} & \textbf{25.69} & \textbf{63.0} && 65.0 \\
        \midrule
        \multicolumn{22}{c}{\textbf{300 points}}\\
        \textbf{GCN2} & 47.0 & 59.0 & 56.17 & 52.0 && - & - & - & - && 40.0 & 39.0 & 44.95 & 40.0 && 38.0 & 43.0 & 45.18 & 40.0 && 44.0 \\
        \textbf{GCN7} & 72.0 & 40.0 & 12.76 & 52.0 && - & - & 0.19 & - && 54.0 & 92.0 & 60.75 & 68.0 && 56.0 & 58.0 & 32.4 & 53.0 && 59.0 \\
        \textbf{CNN-2Conv} & 61.0 & 54.0 & 28.97 & 57.0 && 40.0 & 12.0 & 2.32 & 19.0 && 54.0 & 69.0 & 46.48 & 60.0 && 55.0 & 56.0 & 33.7 & 54.0 && 56.0 \\
        \bottomrule
    \end{tabular}}
     {\footnotesize $^*Acc \Rightarrow$ Accuracy, $Se \Rightarrow$ Sensitivity, $+P \Rightarrow$ Positive Prediction, $FPR \Rightarrow$ False Positive Rate, $F_s \Rightarrow$ F1-score. \newline 
    \textbf{Note:} Values in bold indicate the best performance.}
\end{table*}

\subsection{Experiment 3: Features Aggregation at Graph Vertices}
\label{subsec:experiment_3}

The capability to aggregate feature within graphs, both at vertices and edges, provides an opportunity for incorporating auxiliary and extrinsic data into the graph structure, enhancing the performance of GCNs. This phase assesses the impact of feature aggregation on GCN architectures, focusing on extrinsic graph data, as the core purpose of GCNs is to learn from the inherent structures of the graphs used during training. We did not include the CNN model in this experiment.

For clarity in understanding the experiments and the ensuing results, the aggregated vertex features is categorized as follows:

\begin{itemize}
    \item \textbf{II\_V1}: values from lead II, lead V1, and the timing of each point (03 features).
    \item \textbf{RR}: all data from II\_V1, including the preceding RR interval, and succeeding RR interval (05 features).
    \item \textbf{DifII}: all data from RR, including the difference between values from lead V1 and lead II (06 features).
    \item \textbf{AvgII}: all data from DifII, including the division of lead V1 values by the mean of lead II (07 features).
    \item \textbf{StdII}: all data from AvgII, including the division of lead V1 values by the standard deviation of lead II (08 features).
    \item \textbf{Stats}: all data from StdII, including the statistical measures (entropy, variance, standard deviation, mean, median, 5th percentile, 25th percentile, 75th percentile, 95th percentile, RMS, kurtosis, skewness, zero\_crossings, and mean\_crossings) of the values from lead II (22 features).
\end{itemize}

This categorization provides a comprehensive view of how varying the amount and type of features aggregated at the vertices affects the performance of GCNs in arrhythmia classification.

This experiment's dataset configurations and training settings align with those used in Experiment 1. The key difference in this phase is the division of the dataset into training (DS1) and testing (DS2) sets, adhering to the inter-patient paradigm outlined in Table~\ref{tab:inter_patient} from Section~\ref{sec:database_aami}. A segmentation size of 280 points is employed due to the findings from Experiment 2.

Before segmentation, it is noteworthy that the ECG signals underwent a normalization process using the \textit{z-score} technique. Normalization of a signal is an approach aimed at equalizing its levels to achieve uniformity. In this context, statistical parameters such as mean ($\mu$) and standard deviation ($\sigma$) are used for calculating the \textit{z-score}, forming part of the process\footnote{$z = \frac{x - \mu}{\sigma}$, where $\mu$ is the mean and $\sigma$ is the standard deviation of the ECG signal points.}. The strategic choice of the \textit{z-score} technique is driven by its ability to preserve the distribution of points within the ECG signals themselves, a feature that includes maintaining fiducial points. Given that ECG signals can exhibit a range of peaks with varying magnitudes, applying methods based on distance or neural networks can pose challenges, mainly due to issues related to gradient exploration \citep{shobanadevi2023classification}.

Table~\ref{tab:results_aggregation_vg} summarizes the performance of the GCN2 and GCN7 architectures with various groups of aggregated using the VG method on the test set (DS2). For the GCN2 architecture, the RR and Stats groups exhibited the best outcomes, with average $F_S$ scores of 68\% and 73\%, respectively. In the case of the GCN7 architecture, the AvgII and Stats groups led the performance, both achieving an average $F_S$ of 77\%.

\begin{table*}[!ht]
    \centering
    \caption{Summary of the performance of GCN architectures regarding aggregating features in the graphs using the VG method.}
    \label{tab:results_aggregation_vg} 
    \resizebox{\linewidth}{!}{
    \begin{tabular}{lccccccccccccccccccccc}
        \toprule
        \multirow{2}{*}{\textbf{Features}} & \multicolumn{4}{c}{\textbf{N}} && \multicolumn{4}{c}{\textbf{S}} && \multicolumn{4}{c}{\textbf{V}} && \multicolumn{4}{c}{\textbf{Weighted Average}} && \multirow{2}{*}{$\mathbf{Acc}^*$}\\
        \cmidrule{2-5} \cmidrule{7-10} \cmidrule{12-15} \cmidrule{17-20} 
         & $\mathbf{+P}^*$ & $\mathbf{Se}^*$ & $\mathbf{FPR}^*$ & $\mathbf{F_s}^*$ && $\mathbf{+P}$ & $\mathbf{Se}$ & $\mathbf{FPR}$ & $\mathbf{F_s}$ && $\mathbf{+P}$ & $\mathbf{Se}$ & $\mathbf{FPR}$ & $\mathbf{F_s}$ && $\mathbf{+P}$ & $\mathbf{Se}$ & $\mathbf{FPR}$ & $\mathbf{F_s}$ && \\
        \midrule \midrule
        \multicolumn{22}{c}{\textbf{GCN2}}\\
        \textbf{II\_V1} & 59.0 & 85.0 & 50.92 & 70.0 && - & - & - & - && 68.0 & 66.0 & 16.15 & 67.0 && 51.0 & 62.0 & 29.24 & 55.0 && 62.0\\   
        \textbf{RR} & 74.0 & 94.0 & 28.63 & 83.0 && 52.0 & 2.0 & 0.52 & 5.0 && 77.0 & 91.0 & 14.25 & 83.0 && \textbf{71.0} & \textbf{75.0} & \textbf{18.3} & \textbf{68.0} && 75.0\\
        \textbf{DifII} & 73.0 & 94.0 & 31.07 & 82.0 && 41.0 & 2.0 & 0.79 & 4.0 && 77.0 & 87.0 & 13.63 & 81.0 && 68.0 & 74.0 & 19.28 & 67.0 && 74.0\\
        \textbf{AvgII} & 73.0 & 93.0 & 30.65 & 82.0 && 37.0 & 2.0 & 0.77 & 4.0 && 75.0 & 87.0 & 14.55 & 81.0 && 67.0 & 73.0 & 19.39 & 66.0 && 73.0 \\
        \textbf{StdII} & 73.0 & 92.0 & 29.9 & 81.0 && 48.0 & 1.0 & 0.38 & 3.0 && 74.0 & 88.0 & 16.2 & 80.0 && 68.0 & 73.0 & 19.53 & 66.0 && 73.0\\
        \textbf{Stats} & 75.0 & 94.0 & 27.61 & 84.0 && 50.0 & 12.0 & 2.85 & 19.0 && 85.0 & 92.0 & 8.32 & 88.0 && \textbf{74.0} & \textbf{77.0} & \textbf{16.26} & \textbf{73.0} && 77.0\\
        \midrule
        \multicolumn{22}{c}{\textbf{GCN7}}\\
        \textbf{II\_V1} & 62.0 & 85.0 & 46.31 & 72.0 && 31.0 & 5.0 & 2.89 & 9.0 && 82.0 & 78.0 & 8.51 & 80.0 && 63.0 & 67.0 & 25.06 & 62.0 && 67.0\\   
        \textbf{RR} & 75.0 & 90.0 & 26.72 & 81.0 && 62.0 & 30.0 & 4.44 & 41.0 && 88.0 & 89.0 & 6.31 & 88.0 && 77.0 & 78.0 & 15.47 & 76.0 && 78.0\\
        \textbf{DifII} & 76.0 & 88.0 & 24.07 & 82.0 && 60.0 & 29.0 & 4.68 & 39.0 && 86.0 & 93.0 & 7.73 & 89.0 && 76.0 & 78.0 & 14.76 & 76.0 && 78.0\\
        \textbf{AvgII} & 74.0 & 92.0 & 27.9 & 82.0 && 58.0 & 32.0 & 5.7 & 41.0 && 94.0 & 86.0 & 2.89 & 90.0 && \textbf{78.0} & \textbf{79.0} & \textbf{14.68} & \textbf{77.0} && 79.0\\
        \textbf{StdII} & 76.0 & 89.0 & 25.05 & 82.0 && 58.0 & 30.0 & 5.39 & 40.0 && 89.0 & 91.0 & 5.75 & 90.0 && 77.0 & 78.0 & 15.1 & 76.0 && 78.0\\
        \textbf{Stats} & 75.0 & 92.0 & 26.2 & 83.0 && 62.0 & 27.0 & 4.02 & 38.0 && 92.0 & 93.0 & 4.42 & 92.0 && \textbf{78.0} & \textbf{80.0} & \textbf{14.5} & \textbf{77.0} && 80.0\\
        \bottomrule
    \end{tabular}}
     {\footnotesize $^*Acc \Rightarrow$ Accuracy, $Se \Rightarrow$ Sensitivity, $+P \Rightarrow$ Positive Prediction, $FPR \Rightarrow$ False Positive Rate, $F_s \Rightarrow$ F1-score. \newline 
    \textbf{Note:} Values in bold indicate the best performance.}
\end{table*}

\begin{table*}[!ht]
    \centering
    \caption{Summary of the performance of GCN architectures regarding aggregating features in the graphs using the VVG method.}
    \label{tab:results_aggregation_vvg} 
    \resizebox{\linewidth}{!}{
    \begin{tabular}{lccccccccccccccccccccc}
        \toprule
        \multirow{2}{*}{\textbf{Features}} & \multicolumn{4}{c}{\textbf{N}} && \multicolumn{4}{c}{\textbf{S}} && \multicolumn{4}{c}{\textbf{V}} && \multicolumn{4}{c}{\textbf{Weighted Average}} && \multirow{2}{*}{$\mathbf{Acc}^*$}\\
        \cmidrule{2-5} \cmidrule{7-10} \cmidrule{12-15} \cmidrule{17-20} 
        & $\mathbf{+P}^*$ & $\mathbf{Se}^*$ & $\mathbf{FPR}^*$ & $\mathbf{F_s}^*$ && $\mathbf{+P}$ & $\mathbf{Se}$ & $\mathbf{FPR}$ & $\mathbf{F_s}$ && $\mathbf{+P}$ & $\mathbf{Se}$ & $\mathbf{FPR}$ & $\mathbf{F_s}$ && $\mathbf{+P}$ & $\mathbf{Se}$ & $\mathbf{FPR}$ & $\mathbf{F_s}$ && \\
        \midrule \midrule
        \multicolumn{22}{c}{\textbf{GCN2}}\\
        \textbf{II\_V1} & 58.0 & 80.0 & 49.71 & 67.0 && - & - & - & - && 69.0 & 73.0 & 17.12 & 71.0 && 51.0 & 62.0 & 26.44 & 56.0 && 62.0\\  
        \textbf{RR} & 76.0 & 92.0 & 25.57 & 83.0 && 60.0 & 1.0 & 0.13 & 2.0 && 74.0 & 95.0 & 16.71 & 83.0 && \textbf{72.0} & \textbf{75.0} & \textbf{16.07} & \textbf{67.0} && 75.0\\
        \textbf{DifII} & 75.0 & 92.0 & 26.58 & 83.0 && 36.0 & 1.0 & 0.27 & 1.0 && 75.0 & 93.0 & 16.31 & 83.0 && \textbf{67.0} & \textbf{75.0} & 16.4 & 67.0 && 75.0\\
        \textbf{AvgII} & 75.0 & 93.0 & 27.82 & 83.0 && 32.0 & 1.0 & 0.33 & 1.0 && 75.0 & 91.0 & 15.91 & 82.0 && 66.0 & 74.0 & 16.81 & \textbf{67.0} && 74.0\\
        \textbf{StdII} & 75.0 & 93.0 & 27.25 & 83.0 && 15.0 & 0.0 & 0.22 & 0.0 && 75.0 & 92.0 & 15.96 & 83.0 && 63.0 & \textbf{75.0} & 16.57 & \textbf{67.0} && 75.0\\
        \textbf{Stats} & 76.0 & 93.0 & 25.88 & 84.0 && 52.0 & 13.0 & 2.92 & 21.0 && 84.0 & 93.0 & 9.2 & 88.0 && \textbf{74.0} & \textbf{78.0} & \textbf{14.37} & \textbf{73.0} && 78.0\\
        \midrule
        \multicolumn{22}{c}{\textbf{GCN7}}\\
        \textbf{II\_V1} & 63.0 & 77.0 & 40.44 & 69.0 && 34.0 & 2.0 & 1.02 & 4.0 && 73.0 & 88.0 & 17.0 & 80.0 && 61.0 & 66.0 & 22.64 & 60.0 && 66.0\\
        \textbf{RR} & 75.0 & 91.0 & 27.04 & 82.0 && 56.0 & 29.0 & 5.61 & 39.0 && 89.0 & 86.0 & 5.67 & 87.0 && \textbf{76.0} & 77.0 & 14.25 & 75.0 && 77.0\\
         \textbf{DifII} & 77.0 & 91.0 & 25.41 & 83.0 && 52.0 & 27.0 & 11.49 & 15.0 && 76.0 & 73.0 & 11.79 & 75.0 && 65.0 & 69.0 & 18.09 & 67.0 && 69.0\\
        \textbf{AvgII} & 75.0 & 89.0 & 26.7 & 81.0 && 51.0 & 30.0 & 6.82 & 37.0 && 88.0 & 85.0 & 5.85 & 87.0 && 75.0 & 76.0 & 14.37 & 75.0 && 76.0\\
        \textbf{StdII} & 77.0 & 90.0 & 24.05 & 83.0 && 55.0 & 28.0 & 5.44 & 37.0 && 88.0 & 92.0 & 6.23 & 90.0 && \textbf{76.0} & \textbf{79.0} & \textbf{13.12} & \textbf{76.0} && 79.0\\        
        \textbf{Stats} & 76.0 & 91.0 & 24.6 & 83.0 && 56.0 & 29.0 & 5.5 & 39.0 && 92.0 & 93.0 & 4.22 & 93.0 && \textbf{78.0} & \textbf{80.0} & \textbf{12.74} & \textbf{78.0} && 80.0\\
        \bottomrule
    \end{tabular}}
     {\footnotesize $^*Acc \Rightarrow$ Accuracy, $Se \Rightarrow$ Sensitivity, $+P \Rightarrow$ Positive Prediction, $FPR \Rightarrow$ False Positive Rate, $F_s \Rightarrow$ F1-score. \newline 
    \textbf{Note:} Values in bold indicate the best performance.}
\end{table*}

Meanwhile, Table~\ref{tab:results_aggregation_vvg} presents the performance of the GCN2 and GCN7 architectures according to each features group aggregated through the VVG method. For GCN2, the RR and Stats groups again emerged as the top performers, with average $F_S$ scores of 67\% and 73\%, respectively. Notably, the performance in this architecture is similar across both the VG and VVG methods. For GCN7, the top groups were StdII and Stats, with average $F_S$ scores of 76\% and 78\%, respectively.

The Stats group, which contains the most aggregated features - primarily statistical data from ECG signals totaling 22 features - demonstrated the best performance in both architectures. Conversely, the smaller features groups showed variations in performance across the two architectures, except for II\_V1, whose scores are clearly the smallest. The RR group aggregates five features, whereas the AvgII group comprises seven.

The varying performance of GCNs across different features groups can be attributed to a combination of factors. These include the relevance of the aggregated features, the GCNs' capability to capture each group's distinctive features, the complexity of the GCN architectures, and the specific characteristics of the classified arrhythmias. An additional observation is that in the better-performing groups, the confusion matrices highlighted a tendency of the architectures to classify arrhythmic beats as normal less frequently compared with the other groups. This suggests a more accurate detection of arrhythmias, a crucial aspect of practical ECG analysis.

\subsection{Experiment 4: Swapping training and test sets}
\label{subsec:atividade_4}

In Experiment 4, the MIT-BIH dataset is divided into a training set (DS1) and a testing set (DS2) following the inter-patient paradigm (see Table ~\ref{tab:beat_mti_bih}). It is observed that DS2 has a higher number of class S heartbeats compared to DS1. Given that class S presented challenges in performance with the analyzed architectures, this experiment examines whether reversing the datasets—using DS2 for training and DS1 for testing—can enhance the performance of GCN and CNN architectures in classifying class S.

The dataset configurations and training settings are consistent with those used in Experiment 3. For CNN training, a validation set comprising 10\% of DS2 is used. In the case of GCNs, only the training (DS2) and testing (DS1) sets are used. The GCN architectures incorporated the two most effective features groups identified in Experiment 3 for each architecture under VG and VVG conversion methods.

Table~\ref{tab:results_reverse_vg} presents the performance of the architectures in terms of the dataset reversal between DS1 and DS2 using the VG method. The reversal notably improved the performance in class S for both architectures, with a more significant increase observed in the sensitivity (Se) and $F_S$ metrics, especially in the GCN7 architecture. This enhancement in class S performance positively impacted the overall performance of the architectures, leading to an increase in the average overall $F_S$, ranging from 77\% to 84\%. This suggests that training with a dataset with a higher representation of challenging classes like class S can result in better learning and generalization capabilities for the models.

The GCN7 architecture with the AvgII features group exhibited the best overall performance, achieving an 84\% average $F_S$ score. Comparing this result with the data from Table~\ref{tab:results_aggregation_vg}, the reversal of the datasets led to a performance increase of 9.09\%. Similarly, for the CNN-2Conv architecture, there is an improvement in class S performance and overall effectiveness. Regarding CNN, the sensitivity (Se) and $F_S$ for class S increased by 153.33\% and 193.75\%, respectively, while the general performance showed a 20.31\% rise in the average $F_S$ metric.

Evaluating the the performance of GCN architectures by swapping the DS1 and DS2 datasets using the VVG method, the results presented in Table~\ref{tab:results_reverse_vvg} reveal a notable improvement in the performance of class S concerning the Se and $F_S$ metrics. However, a decline was observed in performance regarding the +P metric. Despite this, the overall $F_S$ scores demonstrated an improvement compared to the outcomes reported in Table~\ref{tab:results_aggregation_vvg}. In the GCN2 architecture, the RR and Stats features groups showed an improvement of 16,42\% and 13.7\%, respectively, while in the GCN7 architecture, the StdII and Stats groups exhibited increases of 9.21\% and 9\%, respectively.

A comparison between the VG and VVG methods, as shown in Tables \ref{tab:results_reverse_vg} and \ref{tab:results_reverse_vvg}, reveals an increase in the average $F_s$ values when employing the VVG mapping for the same feature groups. These findings suggest that, in this context, the inclusion of additional lead features in the VVG representation enhances the overall performance of the studied GCN architectures.

\begin{table*}[!ht]
    \centering
    \caption{Summary of the performance of GCN and CNN architectures regarding the datasets DS1 and DS2 reversal using the VG method.}
    \label{tab:results_reverse_vg} 
    \resizebox{\linewidth}{!}{
    \begin{tabular}{lccccccccccccccccccccc}
        \toprule
        \multirow{2}{*}{\textbf{Features}} & \multicolumn{4}{c}{\textbf{N}} && \multicolumn{4}{c}{\textbf{S}} && \multicolumn{4}{c}{\textbf{V}} && \multicolumn{4}{c}{\textbf{Weighted Average}} && \multirow{2}{*}{$\mathbf{Acc}^*$}\\
        \cmidrule{2-5} \cmidrule{7-10} \cmidrule{12-15} \cmidrule{17-20} 
        & $\mathbf{+P}^*$ & $\mathbf{Se}^*$ & $\mathbf{FPR}^*$ & $\mathbf{F_s}^*$ && $\mathbf{+P}$ & $\mathbf{Se}$ & $\mathbf{FPR}$ & $\mathbf{F_s}$ && $\mathbf{+P}$ & $\mathbf{Se}$ & $\mathbf{FPR}$ & $\mathbf{F_s}$ && $\mathbf{+P}$ & $\mathbf{Se}$ & $\mathbf{FPR}$ & $\mathbf{F_s}$ && \\
        \midrule \midrule
        \multicolumn{22}{c}{\textbf{GCN2}}\\
        \textbf{RR} & 82.0 & 85.0 & 18.66 & 83.0 && 27.0 & 27.0 & 8.33 & 27.0 && 79.0 & 74.0 & 13.57 & 77.0 && 75.0 & 75.0 & 15.54 & 75.0 && 75.0\\
        \textbf{Stats} & 89.0 & 84.0 & 10.38 & 86.0 && 39.0 & 50.0 & 8.72 & 44.0 && 90.0 & 90.0 & 7.18 & 90.0 && \textbf{84.0} & \textbf{83.0} & \textbf{8.91} & \textbf{83.0} && 83.0\\
        \midrule
        \multicolumn{22}{c}{\textbf{GCN7}}\\
        \textbf{AvgII} & 86.0 & 90.0 & 14.22 & 88.0 && 63.0 & 57.0 & 3.74 & 60.0 && 85.0 & 83.0 & 9.64 & 84.0 && \textbf{83.0} & \textbf{84.0} & \textbf{11.3} & \textbf{84.0} && 84.0\\
        \textbf{Stats} & 86.0 & 84.0 & 12.7 & 85.0 && 55.0 & 51.0 & 4.68 & 53.0 && 83.0 & 88.0 & 12.52 & 85.0 && 82.0 & 82.0 & 11.81 & 82.0 && 82.0\\
        \midrule
        \multicolumn{22}{c}{\textbf{CNN-2Conv}}\\
        \textbf{No reversal} & 64.0 & 67.0 & 33.28 & 65.0 && 18.0 & 15.0 & 16.54 & 16.0 && 88.0 & 90.0 & 6.44 & 89.0 && 63.0 & 65.0 & 20.92 & 64.0 && 65.0\\
        \textbf{Reversal} & 86.0 & 73.0 & 11.45 & 79.0 && 62.0 & 38.0 & 2.57 & 47.0 && 74.0 & 95.0 & 23.3 & 83.0 && \textbf{79.0} & \textbf{78.0} & \textbf{15.37} & \textbf{77.0} && 78.0\\
        \bottomrule
    \end{tabular}}
     {\footnotesize $^*Acc \Rightarrow$ Accuracy, $Se \Rightarrow$ Sensitivity, $+P \Rightarrow$ Positive Prediction, $FPR \Rightarrow$ False Positive Rate, $F_s \Rightarrow$ F1-score. \newline 
    \textbf{Note:} Values in bold indicate the best performance.}
\end{table*}

\begin{table*}[!ht]
    \centering
    \caption{Summary of the performance of GCN and CNN architectures regarding the datasets DS1 and DS2 reversal using the VVG method.}
    \label{tab:results_reverse_vvg} 
    \resizebox{\linewidth}{!}{
    \begin{tabular}{lccccccccccccccccccccc}
        \toprule
        \multirow{2}{*}{\textbf{Features}} & \multicolumn{4}{c}{\textbf{N}} && \multicolumn{4}{c}{\textbf{S}} && \multicolumn{4}{c}{\textbf{V}} && \multicolumn{4}{c}{\textbf{Weighted Average}} && \multirow{2}{*}{$\mathbf{Acc}^*$}\\
        \cmidrule{2-5} \cmidrule{7-10} \cmidrule{12-15} \cmidrule{17-20} 
        & $\mathbf{+P}^*$ & $\mathbf{Se}^*$ & $\mathbf{FPR}^*$ & $\mathbf{F_s}^*$ && $\mathbf{+P}$ & $\mathbf{Se}$ & $\mathbf{FPR}$ & $\mathbf{F_s}$ && $\mathbf{+P}$ & $\mathbf{Se}$ & $\mathbf{FPR}$ & $\mathbf{F_s}$ && $\mathbf{+P}$ & $\mathbf{Se}$ & $\mathbf{FPR}$ & $\mathbf{F_s}$ && \\
        \midrule \midrule
        \multicolumn{22}{c}{\textbf{GCN2}}\\
        \textbf{RR} & 83.0 & 95.0 & 18.74 & 88.0 && 24.0 & 11.0 & 3.87 & 15.0 && 81.0 & 79.0 & 12.32 & 80.0 && 76.0 & 80.0 & 13.1 & 78.0 && 80.0\\
        \textbf{Stats} & 87.0 & 89.0 & 13.02 & 88.0 && 34.0 & 32.0 & 6.92 & 33.0 && 89.0 & 88.0 & 7.33 & 89.0 && \textbf{82.0} & \textbf{83.0} & \textbf{9.04} & \textbf{83.0} && 83.0\\
        \midrule
        \multicolumn{22}{c}{\textbf{GCN7}}\\
        \textbf{StdII} & 85.0 & 95.0 & 16.12 & 90.0 && 48.0 & 37.0 & 4.47 & 42.0 && 89.0 & 82.0 & 6.73 & 85.0 && 83.0 & 84.0 & \textbf{9.96} & 83.0 && 84.0\\
        \textbf{Stats} & 87.0 & 92.0 & 12.83 & 90.0 && 52.0 & 48.0 & 4.56 & 50.0 && 91.0 & 87.0 & 17.61 & 89.0 && \textbf{85.0} & \textbf{85.0} & 13.44 & \textbf{85.0} && 85.0\\
        \bottomrule
    \end{tabular}}
     {\footnotesize $^*Acc \Rightarrow$ Accuracy, $Se \Rightarrow$ Sensitivity, $+P \Rightarrow$ Positive Prediction, $FPR \Rightarrow$ False Positive Rate, $F_s \Rightarrow$ F1-score. \newline 
    \textbf{Note:} Values in bold indicate the best performance.}
\end{table*}

\subsection{Experiment 5: Intra-patient Paradigm}
\label{subsec:atividade_5}

In Experiment 5, the intra-patient paradigm is explored in contrast to the inter-patient approach. Unlike the inter-patient paradigm, where there is no overlap of data from the same patient between training and testing sets, the intra-patient approach allows for the presence of heartbeat data from the same patient in both training and testing datasets. This experiment, therefore, serves as a comparative analysis of the architecture performance across these two paradigms.

The dataset configurations and training settings follow those used in Experiment 3, with the distinction that the training and testing sets are randomly determined, allowing for the possibility of the same patient's data appearing in both sets.

Analyzing the results in Table~\ref{tab:results_intra_vg} concerning the GCNs and VG method, the GCN2 architecture showed mixed performance changes with different features groups. The RR group showed a performance decline of 7.35\% in the $F_S$ metric compared with the inter-patient paradigm performance (Table~\ref{tab:results_aggregation_vg}), whereas the Stats group exhibited a performance increase of 2.74\%. In the GCN7 architecture, the AvgII and Stats features groups demonstrated improvements of 2.6\% and 12.98\% in the $F_S$ metric, respectively.

When comparing the two paradigms in the CNN-2Conv architecture, a substantial performance increase is observed in both class S and overall performance. The overall performance increase for the CNN-2Conv, measured by the $F_S$ metric, is 50\%. This significant improvement indicates that the intra-patient paradigm, where data from the same patient can appear in training and testing sets, may lead to better learning and adaptation of the models to the specific characteristics of individual patients' ECG signals.

Regarding the VVG method as shown in Table~\ref{tab:results_intra_vvg}, the GCN2 architecture experienced a general performance decrease in the $F_S$ metric in both the RR and Stats features groups, with declines of 1.51\% and 1.39\%, respectively, compared with the results in Table~\ref{tab:results_aggregation_vvg}. Conversely, in the GCN7 architecture, the StdII and Stats features groups showed improved performance with increases of 13.15\% and 14.10\% in the $F_S$ metric, respectively.

\begin{table*}[!ht]
    \centering
    \caption{Summary of the performance of GCN and CNN architectures regarding the intra-patient paradigm using the VG method.}
    \label{tab:results_intra_vg} 
    \resizebox{\linewidth}{!}{
    \begin{tabular}{lccccccccccccccccccccc}
        \toprule
        \multirow{2}{*}{\textbf{Features}} & \multicolumn{4}{c}{\textbf{N}} && \multicolumn{4}{c}{\textbf{S}} && \multicolumn{4}{c}{\textbf{V}} && \multicolumn{4}{c}{\textbf{Weighted Average}} && \multirow{2}{*}{$\mathbf{Acc}^*$}\\
        \cmidrule{2-5} \cmidrule{7-10} \cmidrule{12-15} \cmidrule{17-20} 
        & $\mathbf{+P}^*$ & $\mathbf{Se}^*$ & $\mathbf{FPR}^*$ & $\mathbf{F_s}^*$ && $\mathbf{+P}$ & $\mathbf{Se}$ & $\mathbf{FPR}$ & $\mathbf{F_s}$ && $\mathbf{+P}$ & $\mathbf{Se}$ & $\mathbf{FPR}$ & $\mathbf{F_s}$ && $\mathbf{+P}$ & $\mathbf{Se}$ & $\mathbf{FPR}$ & $\mathbf{F_s}$ && \\
        \midrule \midrule
        \multicolumn{22}{c}{\textbf{GCN2}}\\
        \textbf{RR} & 76.0 & 87.0 & 24.64 & 81.0 && 40.0 & - & 0.08 & - && 63.0 & 85.0 & 25.8 & 73.0 && 64.0 & 70.0 & 20.27 & 63.0 && 70.0\\
        \textbf{Stats} & 77.0 & 91.0 & 23.75 & 83.0 && 65.0 & 21.0 & 2.7 & 32.0 && 82.0 & 93.0 & 10.61 & 87.0 && \textbf{76.0} & \textbf{78.0} & \textbf{15.21} & \textbf{75.0} && 78.0\\
        \midrule
        \multicolumn{22}{c}{\textbf{GCN7}}\\
        \textbf{AvgII} & 77.0 & 94.0 & 24.3 & 85.0 && 93.0 & 31.0 & 0.56 & 46.0 && 86.0 & 94.0 & 7.62 & 90.0 && 83.0 & 82.0 & 14.03 & 79.0 && 82.0\\
        \textbf{Stats} & 85.0 & 93.0 & 13.88 & 89.0 && 87.0 & 60.0 & 2.15 & 71.0 && 90.0 & 95.0 & 5.54 & 92.0 && \textbf{87.0} & \textbf{87.0} & \textbf{8.77} & \textbf{87.0} && 87.0\\
        \midrule
        \multicolumn{22}{c}{\textbf{CNN-2Conv}}\\
        \textbf{Inter-patient} & 62.0 & 72.0 & 38.07 & 67.0 && 21.0 & 14.0 & 12.57 & 17.0 && 88.0 & 86.0 & 6.1 & 87.0 && 63.0 & 66.0 & 22.27 & 64.0 && 66.0\\
        \textbf{Intra-patient} & 95.0 & 97.0 & 4.11 & 96.0 && 94.0 & 90.0 & 1.4 & 92.0 && 99.0 & 98.0 & 0.69 & 98.0 && \textbf{96.0} & \textbf{96.0} & \textbf{2.42} & \textbf{96.0} && 96.0\\
        \bottomrule
    \end{tabular}}
     {\footnotesize $^*Acc \Rightarrow$ Accuracy, $Se \Rightarrow$ Sensitivity, $+P \Rightarrow$ Positive Prediction, $FPR \Rightarrow$ False Positive Rate, $F_s \Rightarrow$ F1-score. \newline 
    \textbf{Note:} Values in bold indicate the best performance.}
\end{table*}

\begin{table*}[!ht]
    \centering
    \caption{Summary of the performance of GCN and CNN architectures regarding the intra-patient paradigm using the VVG method.}
    \label{tab:results_intra_vvg} 
    \resizebox{\linewidth}{!}{
    \begin{tabular}{lccccccccccccccccccccc}
        \toprule
        \multirow{2}{*}{\textbf{Features}} & \multicolumn{4}{c}{\textbf{N}} && \multicolumn{4}{c}{\textbf{S}} && \multicolumn{4}{c}{\textbf{V}} && \multicolumn{4}{c}{\textbf{Weighted Average}} && \multirow{2}{*}{$\mathbf{Acc}^*$}\\
        \cmidrule{2-5} \cmidrule{7-10} \cmidrule{12-15} \cmidrule{17-20} 
        & $\mathbf{+P}^*$ & $\mathbf{Se}^*$ & $\mathbf{FPR}^*$ & $\mathbf{F_s}^*$ && $\mathbf{+P}$ & $\mathbf{Se}$ & $\mathbf{FPR}$ & $\mathbf{F_s}$ && $\mathbf{+P}$ & $\mathbf{Se}$ & $\mathbf{FPR}$ & $\mathbf{F_s}$ && $\mathbf{+P}$ & $\mathbf{Se}$ & $\mathbf{FPR}$ & $\mathbf{F_s}$ && \\
        \midrule \midrule
        \multicolumn{22}{c}{\textbf{GCN2}}\\
        \textbf{RR} & 76.0 & 89.0 & 24.07 & 82.0 && 89.0 & 3.0 & 0.09 & 6.0 && 68.0 & 91.0 & 21.77 & 78.0 && \textbf{76.0} & 73.0 & 16.99 & 66.0 && 73.0\\
        \textbf{Stats} & 77.0 & 93.0 & 24.03 & 84.0 && 68.0 & 12.0 & 1.35 & 21.0 && 78.0 & 94.0 & 13.21 & 85.0 && \textbf{76.0} & \textbf{77.0} & \textbf{14.55} & \textbf{72.0} && 77.0\\
        \midrule
        \multicolumn{22}{c}{\textbf{GCN7}}\\
        \textbf{StdII} & 85.0 & 91.0 & 13.64 & 88.0 && 83.0 & 63.0 & 3.04 & 72.0 && 89.0 & 93.0 & 5.96 & 91.0 && 86.0 & 86.0 & 8.18 & 86.0 && 86.0\\
        \textbf{Stats} & 88.0 & 93.0 & 11.49 & 90.0 && 84.0 & 70.0 & 3.27 & 76.0 && 94.0 & 94.0 & 3.12 & 94.0 && \textbf{89.0} & \textbf{89.0} & \textbf{6.43} & \textbf{89.0} && 89.0\\
        \bottomrule
    \end{tabular}}
     {\footnotesize $^*Acc \Rightarrow$ Accuracy, $Se \Rightarrow$ Sensitivity, $+P \Rightarrow$ Positive Prediction, $FPR \Rightarrow$ False Positive Rate, $F_s \Rightarrow$ F1-score. \newline 
    \textbf{Note:} Values in bold indicate the best performance.}
\end{table*}

This variation in performance between the GCN2 and GCN7 architectures under the intra-patient paradigm using the VVG method suggests that different features groups and GCN architectures respond uniquely to the challenges posed by this paradigm. Specifically, while some architectures and features groups may struggle to adapt to the overlapping patient data in the intra-patient setting, others, such as GCN7 with the StdII and Stats groups, appear to thrive, showing significant performance improvements. This observation highlights the importance of selecting the appropriate combination of architecture and features groups to optimize performance in a clinical or experimental setting.

Overall, the results underscore that the intra-patient paradigm generally yields better outcomes than the inter-patient paradigm in the performance of the examined architectures despite not representing a scenario closer to real-world conditions where data from a new patient are not used during model training.

Throughout the experiments, we observe that the VVG method exhibits high computational complexity in both time and space due to using two leads for mapping ECG signals. However, space complexity remains challenging in mapping ECG signals and processing the generated graphs during training and testing. We initially applied data subsampling to mitigate this issue and explore other solutions, allowing us to leverage the full potential of graph convolutional networks. This exploration might involve investigating alternative data representations or optimizing graph processing algorithms to enhance the efficiency of space utilization while maintaining the integrity and effectiveness of the analysis.

\subsection{Experiment 6: Comparison of the Proposed Method}
\label{subsec:experiment_6}

In this final experiment, we conduct a comparative analysis between the method proposed in this work and the study by \citet{garcia2017inter}, which employed graph modeling and an SVM classifier. The rationale for selecting this study for comparison stems from its status as a significant baseline reference in this work and its methodological parallels with the approach presented herein. Modifications have been made to the original study by \citet{garcia2017inter} to enable direct comparison. Notably, we included a 10\% subsampling of class N heartbeats by selecting only the tenth beat in every sequence of ten. Furthermore, the same datasets, DS1.1 and DS1.2, as detailed in Table~\ref{tab:DS1_inter_patient}, are employed for tuning the SVM model parameters. The comparison extended across three distinct scenarios:

\begin{itemize}
    \item \citet{garcia2017inter}:
    \begin{itemize}
        \item \textbf{Scenario 1}: Features extracted from the graph network based on complex networks;
        \item \textbf{Scenario 2}: Features extracted from the graph network based on complex networks along with RR interval features;
        \item \textbf{Scenario 3}: Features extracted from the graph network based on complex networks, RR interval features, and statistical characteristics of the graph network;
    \end{itemize} 
    \item \textbf{Proposed Method}:
    \begin{itemize}
        \item \textbf{Scenario 1}: Experiment 3 of the GCN7 architecture with features from the II\_V1 group and the VG method;
        \item \textbf{Scenario 2}: Experiment 3 of the GCN7 architecture with features from the RR group and the VG method;
        \item \textbf{Scenario 3}: Experiment 3 of the GCN7 architecture with features from the Stats group and the VG method;
    \end{itemize}  
\end{itemize}

\begin{table*}[!ht]
    \centering
    \caption{Comparison of the proposed method with the work of \citet{garcia2017inter}.}
    \label{tab:comparativo_garcia} 
    \resizebox{\linewidth}{!}{
    \begin{tabular}{cccccccccccccccccc}
        \toprule
        \multirow{2}{*}{\textbf{Scenarios}} & \multirow{2}{*}{\textbf{Work}} & \multicolumn{4}{c}{\textbf{N}} && \multicolumn{4}{c}{\textbf{S}} && \multicolumn{4}{c}{\textbf{V}} && \multirow{2}{*}{$\mathbf{Acc}^*$} \\ 
        \cmidrule{3-6} \cmidrule{8-11} \cmidrule{13-16}
        && $\mathbf{+P}^*$ & $\mathbf{Se}^*$ & $\mathbf{FPR}^*$ & $\mathbf{F_s}^*$ && $\mathbf{+P}$ & $\mathbf{Se}$ & $\mathbf{FPR}$ & $\mathbf{F_s}$ && $\mathbf{+P}$ & $\mathbf{Se}$ & $\mathbf{FPR}$ & $\mathbf{F_s}$ \\
        \midrule \midrule        
        \multirow{2}{*}{\textbf{1}} & Garcia et al. \citep{garcia2017inter} & \textbf{79.60} & 74.2 & 60.83 & \textbf{76.80} && 0.40 & 0.20 & 7.90 & 0.30 && 39.40 & 71.9 & 15.13 & 50.90 && 65.20\\
        & Proposed method & 62.0 & \textbf{85.0} & \textbf{46.31} & 71.47 && \textbf{31.0} & \textbf{5.0} & \textbf{2.89} & \textbf{8.61} && \textbf{82.0} & \textbf{78.0} & \textbf{8.51} & \textbf{79.95} && \textbf{67.0}\\
        \midrule 
        \multirow{2}{*}{\textbf{2}} & Garcia et al. \citep{garcia2017inter} & \textbf{86.60} & 87.40 & 43.36 & \textbf{87.0} && 21.10 & 2.80 & \textbf{1.38} & 4.90 && 46.80 & 83.90 & 13.05 & 60.10 && 77.0\\ 
        & Proposed method & 75.0 & \textbf{90.0} & \textbf{26.72} & 81.81 && \textbf{62.0} & \textbf{30.0} & 4.44 & \textbf{40.43} && \textbf{88.0} & \textbf{89.0} & \textbf{6.31} & \textbf{88.50} && \textbf{78.0}\\
        \midrule 
        \multirow{2}{*}{\textbf{3}} & Garcia et al. \citep{garcia2017inter} & \textbf{90.20} & \textbf{95.30} & 48.90 & \textbf{92.70} && 55.30 & 17.70 & \textbf{1.37} & 26.80 && 68.20 & 78.40 & \textbf{3.48} & 72.90 && \textbf{87.0}\\ 
        & Proposed method & 78.0 & 92.0 & \textbf{26.20} & 84.82 && \textbf{62.0} & \textbf{27.0} & 4.02 & \textbf{37.62} && \textbf{92.0} & \textbf{93.0} & 4.42 & \textbf{92.49} && 80.0\\
        \bottomrule
        \end{tabular}}
        {\footnotesize $^*Acc \Rightarrow$ Accuracy, $Se \Rightarrow$ Sensitivity, $+P \Rightarrow$ Positive Prediction, $FPR \Rightarrow$ False Positive Rate, $F_s \Rightarrow$ F1-score. \newline 
        \textbf{Note:} Values in bold indicate the best performance.}
\end{table*}

The comparative results in Table~\ref{tab:comparativo_garcia} highlight the efficacy of our introduced method when contrasted with the findings of \citet{garcia2017inter}. In both Scenario 1 and Scenario 2, the proposed method delivered improved results for minority classes, specifically the arrhythmic categories, but encountered challenges with class N, where it displayed a modest reduction in performance in Positive Prediction (+P) and $F_s$ metrics. A similar trend is evident in Scenario 3, with class N exhibiting a decline in performance concerning the +P, Sensitivity (Se), and $F_s$ metrics. However, it is essential to acknowledge that in Scenario 3, the statistical features used by \citet{garcia2017inter} differ from the features used in the proposed method. Despite this disparity, a comparative analysis enables an approximate evaluation of the proposed scenario.

A key observation is that the proposed method demonstrates superior performance in the arrhythmic classes (S and V) compared with the normal class (N), underscoring its effectiveness in differentiating between normal and arrhythmic heartbeats. This differentiation is essential for minimizing false predictions, a significant concern in clinical applications that impacts the method's reliability.

\subsection{Experimental Decisions }
\label{subsec:decisions_experimental}

Acknowledging the stochastic nature of neural network training, we conducted each experiment only once. Preliminary tests guided this decision, where we ran selected architectures ten times each, revealing minimal variation and deviation in the outcomes, as detailed in Table~\ref{tab:repeated_experiment}. These initial experiments, focusing on the GCN2 and GCN7 architectures with the Stats features set and employing the VG method, demonstrated significant consistency in standard deviation and variance across evaluated metrics, even with the introduction of randomness through varied training seeds.\footnote{We introduced randomness via the \textbf{torch.manual\_seed( random.randint(1,100000)} command, utilizing the \textit{torch} and \textit{random} libraries to generate a random number between 1 and 100000 for each execution.} Given the observed stability and the high computational cost of multiple runs, we concluded that a single execution would suffice, ensuring efficient resource use while maintaining confidence in the reliability and reproducibility of our results.
\begin{table*}[!ht]
    \centering
    \caption{Results of experimental decisions.}
    \label{tab:repeated_experiment}
    \begin{tabular}{lcccccc}
        \toprule 
        $\mathbf{\# Execution}$ & $\mathbf{Acc}^*$ & $\mathbf{+P}^*$ & $\mathbf{Se}^*$ & $\mathbf{FPR}^*$ & $\textbf{F}_s^*$ & $\mathbf{Time (s)}$\\
        \midrule
        \multicolumn{7}{c}{\textbf{Experiment 1: GCN2\_Stats}}\\
        \midrule        
        \#1 & 77.56 & 74.62 & 77.56 & 16.33 & 72.17 & 5408\\
        \#2 & 77.33 & 73.41 & 77.33 & 16.07 & 72.78 & 5435\\
        \#3 & 78.05 & 75.50 & 78.04 & 16.21 & 72.84 & 5528\\
        \#4 & 77.40 & 73.84 & 77.40 & 16.41 & 72.56 & 5521\\
        \#5 & 77.70 & 74.29 & 77.70 & 16.04 & 72.96 & 5528\\
        \#6 & 77.80 & 74.41 & 77.80 & 15.98 & 73.05 & 5555\\
        \#7 & 77.37 & 73.64 & 77.37 & 16.23 & 72.32 & 5567\\
        \#8 & 77.48 & 73.50 & 77.48 & 15.95 & 72.90 & 5578\\
        \#9 & 76.88 & 72.77 & 76.88 & 16.20 & 72.75 & 5598\\
        \#10 & 77.48 & 73.71 & 77.48 & 16.22 & 72.48 & 5617\\
        \hdashline
        \textbf{Average} & 77.50 & 73.97 & 77.51 & 16.16 & 72.68 & 5533\\
        \textbf{Standard Deviation} & 0.003 & 0.007 & 0.003 & 0.14 & 0.003 & 63.53\\
        \textbf{Variance} & {8.8e-6} & {5.23e-5} & {8.8e-6} & {0.02} & {7.5e-6} & {4.04e3}\\
        \midrule\midrule
        \multicolumn{7}{c}{\textbf{Experiment 2: GCN7\_Stats}}\\ 
        \midrule        
        \#1 & 78.15 & 76.33 & 78.15 & 15.17 & 75.52 & 8124\\
        \#2 & 79.42 & 78.32 & 79.42 & 15.16 & 76.95 & 8174\\
        \#3 & 79.56 & 78.20 & 79.58 & 14.30 & 77.74 & 8290\\
        \#4 & 80.06 & 79.66 & 80.06 & 15.0 & 77.26 & 8337\\
        \#5 & 78.86 & 77.21 & 78.86 & 14.50 & 76.83 & 8390\\
        \#6 & 79.75 & 78.21 & 79.75 & 14.09 & 77.56 & 8370\\
        \#7 & 79.04 & 77.66 & 79.04 & 15.17 & 76.37 & 8378\\
        \#8 & 78.62 & 77.00 & 78.62 & 15.19 & 75.08 & 8395\\
        \#9 & 78.37 & 77.33 & 78.37 & 15.25 & 77.76 & 8462\\
        \#10 & 79.64 & 78.49 & 79.64 & 14.65 & 77.56 & 8437\\
        \hdashline
        \textbf{Average} & 79.15 & 77.84 & 79.15 & 14.84 & 76.66 & 8335\\
        \textbf{Standard Deviation} & 0.006 & 0.009 & 0.006 & 0.405 & 0.009 & 104.26\\
        \textbf{Variance} & {3.66e-5} & {7.90e-5} & {3.66e-5} & {0.16} & {8.0e-5} & {1.09e4} \\
        \bottomrule
    \end{tabular}
    
    \vspace{0.1cm}
    {\footnotesize $^*$ $Acc \Rightarrow$ Accuracy, $Se \Rightarrow$ Sensitivity, $+P \Rightarrow$ Positive Prediction, $FPR \Rightarrow$ False Positive Rate, $F_s \Rightarrow$ F1-score.}
\end{table*}

\section{Conclusion}
\label{sec-conclusion}

This study proposed a method for classifying arrhythmias in ECG signals by mapping them into graphs and classifying them using Graph Convolutional Networks (GCNs), following the inter-patient paradigm and AAMI standards. The central research question is whether graph representations of ECG signals, through VG and VVG methods, could enhance arrhythmia classification performance using GCNs.

The findings indicated that simpler GCN architectures yielded better results than more complex ones, suggesting that simplicity in GCN structures can more effectively capture essential data characteristics and avoid unnecessary noise. The results presented in Section~\ref{subsec:experiment_1} showed that GCN2 and GCN7 outperformed in most metrics, obtaining 41\% and 54\% in average $F_s$ metric, respectively.

Simpler architectures are computationally more efficient, a necessary factor in resource-constrained scenarios. Including extrinsic features in the graphs, such as features describing the morphology of the ECG signal, may enhance the performance of models trained using data derived from the VG or VVG methods. Both VG and VVG are promise, as shown by the results in Experiment 3 on Section~\ref{subsec:experiment_3}. Considering the best performances, the GCN architectures achieved average $F_s$ metric scores of 77\% and 78\% with the VG and VVG methods, respectively.

The classification of S class remained challenging, particularly under the interpatient paradigm, even when reversing the DS1 and DS2 datasets. Conversely, the intrapatient paradigm yielded better results, although it does not fully mirror real-world scenarios. The CNN architecture achieved a score of 96\%, while the GCN7 architecture with VVG reached 89\% in $F_s$. These findings indicate that classifying arrhythmias in ECG signals using GCNs with VG and VVG for signal graph mapping is feasible, offering the added benefit of bypassing the need for preprocessing or denoising the signals. Nonetheless, there is significant potential for improvement, warranting further research to fully harness the capabilities of GCNs.

A significant limitation encountered in developing this method is the computational complexity of the VVG method in mapping ECG signals and training the GCNs, which is related to processing time and required space. However, optimizations and development alternatives can be explored to mitigate this limitation.

\section*{Statements and Declarations}

\bmhead{Acknowledgments}

This study was financed in part by the Coordenação de Aperfeiçoamento de Pessoal de Nível Superior – Brasil (CAPES) - Finance Code 001, by Fundacão de Amparo à Pesquisa do Estado de Minas Gerais (FAPEMIG, grants APQ-01518-21, APQ-01647-22), and by Conselho Nacional de Desenvolvimento Científico e Tecnológico (CNPq, grants 307151/2022-0, 308400/2022-4).

\bmhead{Competing Interests}

The authors have no relevant financial or non-financial interests to disclose.

\bmhead{Author Contributions}

\textbf{Rafael F. Oliveira:} Investigation, Methodology, Software, Validation, Visualization, Writing – original draft; \textbf{Gladston J. P. Moreira:} Resources, Writing – review \& editing; \textbf{Eduardo J. S. Luz:} Conceptualization, Data curation, Formal analysis, Funding acquisition, Investigation, Methodology, Project administration, Resources, Supervision, Validation, Writing – review \& editing; \textbf{Vander L. S. Freitas:} Conceptualization, Investigation, Methodology, Project administration, Supervision, Software, Validation, Writing – review \& editing.

\bmhead{Data Availability}

The datasets generated during and/or analysed during the current study are available in the following repository: \url{https://github.com/raffoliveira/VG_for_arrhythmia_classification_with_GCN}

\bibliography{sn-bibliography}


\begin{thebibliography}{47}
\ifx \bisbn   \undefined \def \bisbn  #1{ISBN #1}\fi
\ifx \binits  \undefined \def \binits#1{#1}\fi
\ifx \bauthor  \undefined \def \bauthor#1{#1}\fi
\ifx \batitle  \undefined \def \batitle#1{#1}\fi
\ifx \bjtitle  \undefined \def \bjtitle#1{#1}\fi
\ifx \bvolume  \undefined \def \bvolume#1{\textbf{#1}}\fi
\ifx \byear  \undefined \def \byear#1{#1}\fi
\ifx \bissue  \undefined \def \bissue#1{#1}\fi
\ifx \bfpage  \undefined \def \bfpage#1{#1}\fi
\ifx \blpage  \undefined \def \blpage #1{#1}\fi
\ifx \burl  \undefined \def \burl#1{\textsf{#1}}\fi
\ifx \doiurl  \undefined \def \doiurl#1{\url{https://doi.org/#1}}\fi
\ifx \betal  \undefined \def \betal{\textit{et al.}}\fi
\ifx \binstitute  \undefined \def \binstitute#1{#1}\fi
\ifx \binstitutionaled  \undefined \def \binstitutionaled#1{#1}\fi
\ifx \bctitle  \undefined \def \bctitle#1{#1}\fi
\ifx \beditor  \undefined \def \beditor#1{#1}\fi
\ifx \bpublisher  \undefined \def \bpublisher#1{#1}\fi
\ifx \bbtitle  \undefined \def \bbtitle#1{#1}\fi
\ifx \bedition  \undefined \def \bedition#1{#1}\fi
\ifx \bseriesno  \undefined \def \bseriesno#1{#1}\fi
\ifx \blocation  \undefined \def \blocation#1{#1}\fi
\ifx \bsertitle  \undefined \def \bsertitle#1{#1}\fi
\ifx \bsnm \undefined \def \bsnm#1{#1}\fi
\ifx \bsuffix \undefined \def \bsuffix#1{#1}\fi
\ifx \bparticle \undefined \def \bparticle#1{#1}\fi
\ifx \barticle \undefined \def \barticle#1{#1}\fi
\bibcommenthead
\ifx \bconfdate \undefined \def \bconfdate #1{#1}\fi
\ifx \botherref \undefined \def \botherref #1{#1}\fi
\ifx \url \undefined \def \url#1{\textsf{#1}}\fi
\ifx \bchapter \undefined \def \bchapter#1{#1}\fi
\ifx \bbook \undefined \def \bbook#1{#1}\fi
\ifx \bcomment \undefined \def \bcomment#1{#1}\fi
\ifx \oauthor \undefined \def \oauthor#1{#1}\fi
\ifx \citeauthoryear \undefined \def \citeauthoryear#1{#1}\fi
\ifx \endbibitem  \undefined \def \endbibitem {}\fi
\ifx \bconflocation  \undefined \def \bconflocation#1{#1}\fi
\ifx \arxivurl  \undefined \def \arxivurl#1{\textsf{#1}}\fi
\csname PreBibitemsHook\endcsname

\bibitem[\protect\citeauthoryear{{ANSI/AAMI}}{2008}]{aami:2008}
\begin{botherref}
\oauthor{\bsnm{{ANSI/AAMI}}}:
Testing and Reporting Performance Results of Cardiac Rhythm and {ST} Segment Measurement Algorithms.
American National Standards Institute, Inc. (ANSI), Association for the Advancement of Medical Instrumentation (AAMI).
{ANSI/AAMI/ISO} EC57, 1998-(R)2008
(2008)
\end{botherref}
\endbibitem

\bibitem[\protect\citeauthoryear{Barab{\'a}si}{2013}]{barabasi2013network}
\begin{barticle}
\bauthor{\bsnm{Barab{\'a}si}, \binits{A.-L.}}:
\batitle{Network science}.
\bjtitle{Philosophical Transactions of the Royal Society A: Mathematical, Physical and Engineering Sciences}
\bvolume{371}(\bissue{1987}),
\bfpage{20120375}
(\byear{2013})
\end{barticle}
\endbibitem

\bibitem[\protect\citeauthoryear{Cheng and Dong}{2017}]{cheng2017life}
\begin{barticle}
\bauthor{\bsnm{Cheng}, \binits{P.}},
\bauthor{\bsnm{Dong}, \binits{X.}}:
\batitle{Life-threatening ventricular arrhythmia detection with personalized features}.
\bjtitle{IEEE access}
\bvolume{5},
\bfpage{14195}--\blpage{14203}
(\byear{2017})
\end{barticle}
\endbibitem

\bibitem[\protect\citeauthoryear{Cohen}{2019}]{cohen2019biomedical}
\begin{botherref}
\oauthor{\bsnm{Cohen}, \binits{A.}}:
Biomedical signal processing: Compression and automatic recognition
\textbf{2}
(2019)
\end{botherref}
\endbibitem

\bibitem[\protect\citeauthoryear{Cao et~al.}{2023}]{cao2023ecg}
\begin{bchapter}
\bauthor{\bsnm{Cao}, \binits{M.}},
\bauthor{\bsnm{Zhao}, \binits{T.}},
\bauthor{\bsnm{Li}, \binits{Y.}},
\bauthor{\bsnm{Zhang}, \binits{W.}},
\bauthor{\bsnm{Benharash}, \binits{P.}},
\bauthor{\bsnm{Ramezani}, \binits{R.}}:
\bctitle{Ecg heartbeat classification using deep transfer learning with convolutional neural network and stft technique}.
In: \bbtitle{Journal of Physics: Conference Series},
vol. \bseriesno{2547},
p. \bfpage{012031}
(\byear{2023}).
\bcomment{IOP Publishing}
\end{bchapter}
\endbibitem

\bibitem[\protect\citeauthoryear{De~Chazal et~al.}{2004}]{de2004automatic}
\begin{barticle}
\bauthor{\bsnm{De~Chazal}, \binits{P.}},
\bauthor{\bsnm{O'Dwyer}, \binits{M.}},
\bauthor{\bsnm{Reilly}, \binits{R.B.}}:
\batitle{Automatic classification of heartbeats using ecg morphology and heartbeat interval features}.
\bjtitle{IEEE transactions on biomedical engineering}
\bvolume{51}(\bissue{7}),
\bfpage{1196}--\blpage{1206}
(\byear{2004})
\end{barticle}
\endbibitem

\bibitem[\protect\citeauthoryear{Duong et~al.}{2023}]{DUONG2023120107}
\begin{barticle}
\bauthor{\bsnm{Duong}, \binits{L.T.}},
\bauthor{\bsnm{Doan}, \binits{T.T.H.}},
\bauthor{\bsnm{Chu}, \binits{C.Q.}},
\bauthor{\bsnm{Nguyen}, \binits{P.T.}}:
\batitle{Fusion of edge detection and graph neural networks to classifying electrocardiogram signals}.
\bjtitle{Expert Systems with Applications}
\bvolume{225},
\bfpage{120107}
(\byear{2023})
\doiurl{10.1016/j.eswa.2023.120107}
\end{barticle}
\endbibitem

\bibitem[\protect\citeauthoryear{Donges et~al.}{2011}]{donges2011nonlinear}
\begin{barticle}
\bauthor{\bsnm{Donges}, \binits{J.F.}},
\bauthor{\bsnm{Donner}, \binits{R.V.}},
\bauthor{\bsnm{Trauth}, \binits{M.H.}},
\bauthor{\bsnm{Marwan}, \binits{N.}},
\bauthor{\bsnm{Schellnhuber}, \binits{H.-J.}},
\bauthor{\bsnm{Kurths}, \binits{J.}}:
\batitle{Nonlinear detection of paleoclimate-variability transitions possibly related to human evolution}.
\bjtitle{Proceedings of the National Academy of Sciences}
\bvolume{108}(\bissue{51}),
\bfpage{20422}--\blpage{20427}
(\byear{2011})
\end{barticle}
\endbibitem

\bibitem[\protect\citeauthoryear{Donner et~al.}{2010}]{donner2010recurrence}
\begin{barticle}
\bauthor{\bsnm{Donner}, \binits{R.V.}},
\bauthor{\bsnm{Zou}, \binits{Y.}},
\bauthor{\bsnm{Donges}, \binits{J.F.}},
\bauthor{\bsnm{Marwan}, \binits{N.}},
\bauthor{\bsnm{Kurths}, \binits{J.}}:
\batitle{Recurrence networks—a novel paradigm for nonlinear time series analysis}.
\bjtitle{New Journal of Physics}
\bvolume{12}(\bissue{3}),
\bfpage{033025}
(\byear{2010})
\end{barticle}
\endbibitem

\bibitem[\protect\citeauthoryear{Essa and Xie}{2021}]{essa2021ensemble}
\begin{barticle}
\bauthor{\bsnm{Essa}, \binits{E.}},
\bauthor{\bsnm{Xie}, \binits{X.}}:
\batitle{An ensemble of deep learning-based multi-model for ecg heartbeats arrhythmia classification}.
\bjtitle{IEEE Access}
\bvolume{9},
\bfpage{103452}--\blpage{103464}
(\byear{2021})
\end{barticle}
\endbibitem

\bibitem[\protect\citeauthoryear{{Freitas} et~al.}{2019}]{Freitas_et_al_2019}
\begin{barticle}
\bauthor{\bsnm{{Freitas}}, \binits{V.L.S.}},
\bauthor{\bsnm{{Lacerda}}, \binits{J.C.}},
\bauthor{\bsnm{{Macau}}, \binits{E.E.N.}}:
\batitle{{Complex Networks Approach for Dynamical Characterization of Nonlinear Systems}}.
\bjtitle{International Journal of Bifurcation and Chaos}
\bvolume{29}(\bissue{13}),
\bfpage{1950188}--\blpage{512}
(\byear{2019})
\doiurl{10.1142/S0218127419501888}
\end{barticle}
\endbibitem

\bibitem[\protect\citeauthoryear{Gai}{2022}]{gai2022ecg}
\begin{botherref}
\oauthor{\bsnm{Gai}, \binits{N.D.}}:
Ecg beat classification using machine learning and pre-trained convolutional neural networks.
arXiv preprint arXiv:2207.06408
(2022)
\end{botherref}
\endbibitem

\bibitem[\protect\citeauthoryear{Gotoda et~al.}{2017}]{gotoda2017characterization}
\begin{barticle}
\bauthor{\bsnm{Gotoda}, \binits{H.}},
\bauthor{\bsnm{Kinugawa}, \binits{H.}},
\bauthor{\bsnm{Tsujimoto}, \binits{R.}},
\bauthor{\bsnm{Domen}, \binits{S.}},
\bauthor{\bsnm{Okuno}, \binits{Y.}}:
\batitle{Characterization of combustion dynamics, detection, and prevention of an unstable combustion state based on a complex-network theory}.
\bjtitle{Physical Review Applied}
\bvolume{7}(\bissue{4}),
\bfpage{044027}
(\byear{2017})
\end{barticle}
\endbibitem

\bibitem[\protect\citeauthoryear{Garcia et~al.}{2017}]{garcia2017inter}
\begin{barticle}
\bauthor{\bsnm{Garcia}, \binits{G.}},
\bauthor{\bsnm{Moreira}, \binits{G.}},
\bauthor{\bsnm{Menotti}, \binits{D.}},
\bauthor{\bsnm{Luz}, \binits{E.}}:
\batitle{Inter-patient ecg heartbeat classification with temporal vcg optimized by pso}.
\bjtitle{Scientific reports}
\bvolume{7}(\bissue{1}),
\bfpage{1}--\blpage{11}
(\byear{2017})
\end{barticle}
\endbibitem

\bibitem[\protect\citeauthoryear{Hannun et~al.}{2019}]{hannun2019cardiologist}
\begin{barticle}
\bauthor{\bsnm{Hannun}, \binits{A.Y.}},
\bauthor{\bsnm{Rajpurkar}, \binits{P.}},
\bauthor{\bsnm{Haghpanahi}, \binits{M.}},
\bauthor{\bsnm{Tison}, \binits{G.H.}},
\bauthor{\bsnm{Bourn}, \binits{C.}},
\bauthor{\bsnm{Turakhia}, \binits{M.P.}},
\bauthor{\bsnm{Ng}, \binits{A.Y.}}:
\batitle{Cardiologist-level arrhythmia detection and classification in ambulatory electrocardiograms using a deep neural network}.
\bjtitle{Nature medicine}
\bvolume{25}(\bissue{1}),
\bfpage{65}
(\byear{2019})
\end{barticle}
\endbibitem

\bibitem[\protect\citeauthoryear{Hamilton et~al.}{2017}]{hamilton2017inductive}
\begin{bchapter}
\bauthor{\bsnm{Hamilton}, \binits{W.L.}},
\bauthor{\bsnm{Ying}, \binits{R.}},
\bauthor{\bsnm{Leskovec}, \binits{J.}}:
\bctitle{Inductive representation learning on large graphs}.
In: \bbtitle{Proceedings of the 31st International Conference on Neural Information Processing Systems},
pp. \bfpage{1025}--\blpage{1035}
(\byear{2017})
\end{bchapter}
\endbibitem

\bibitem[\protect\citeauthoryear{Kojima et~al.}{2020}]{kojima2020kgcn}
\begin{barticle}
\bauthor{\bsnm{Kojima}, \binits{R.}},
\bauthor{\bsnm{Ishida}, \binits{S.}},
\bauthor{\bsnm{Ohta}, \binits{M.}},
\bauthor{\bsnm{Iwata}, \binits{H.}},
\bauthor{\bsnm{Honma}, \binits{T.}},
\bauthor{\bsnm{Okuno}, \binits{Y.}}:
\batitle{kgcn: a graph-based deep learning framework for chemical structures}.
\bjtitle{Journal of Cheminformatics}
\bvolume{12}(\bissue{1}),
\bfpage{1}--\blpage{10}
(\byear{2020})
\end{barticle}
\endbibitem

\bibitem[\protect\citeauthoryear{Kipf and Welling}{2016}]{kipf2016semi}
\begin{botherref}
\oauthor{\bsnm{Kipf}, \binits{T.N.}},
\oauthor{\bsnm{Welling}, \binits{M.}}:
Semi-supervised classification with graph convolutional networks.
arXiv:1609.02907
(2016)
\end{botherref}
\endbibitem

\bibitem[\protect\citeauthoryear{LeCun et~al.}{1998}]{lecun1998gradient}
\begin{barticle}
\bauthor{\bsnm{LeCun}, \binits{Y.}},
\bauthor{\bsnm{Bottou}, \binits{L.}},
\bauthor{\bsnm{Bengio}, \binits{Y.}},
\bauthor{\bsnm{Haffner}, \binits{P.}}:
\batitle{Gradient-based learning applied to document recognition}.
\bjtitle{Proceedings of the IEEE}
\bvolume{86}(\bissue{11}),
\bfpage{2278}--\blpage{2324}
(\byear{1998})
\end{barticle}
\endbibitem

\bibitem[\protect\citeauthoryear{Lacasa et~al.}{2008}]{lacasa2008time}
\begin{barticle}
\bauthor{\bsnm{Lacasa}, \binits{L.}},
\bauthor{\bsnm{Luque}, \binits{B.}},
\bauthor{\bsnm{Ballesteros}, \binits{F.}},
\bauthor{\bsnm{Luque}, \binits{J.}},
\bauthor{\bsnm{Nuno}, \binits{J.C.}}:
\batitle{From time series to complex networks: The visibility graph}.
\bjtitle{Proceedings of the National Academy of Sciences}
\bvolume{105}(\bissue{13}),
\bfpage{4972}--\blpage{4975}
(\byear{2008})
\end{barticle}
\endbibitem

\bibitem[\protect\citeauthoryear{Luque et~al.}{2009}]{luque2009horizontal}
\begin{barticle}
\bauthor{\bsnm{Luque}, \binits{B.}},
\bauthor{\bsnm{Lacasa}, \binits{L.}},
\bauthor{\bsnm{Ballesteros}, \binits{F.}},
\bauthor{\bsnm{Luque}, \binits{J.}}:
\batitle{Horizontal visibility graphs: Exact results for random time series}.
\bjtitle{Physical Review E}
\bvolume{80}(\bissue{4}),
\bfpage{046103}
(\byear{2009})
\end{barticle}
\endbibitem

\bibitem[\protect\citeauthoryear{Luz and Menotti}{2011}]{luz2011choice}
\begin{bchapter}
\bauthor{\bsnm{Luz}, \binits{E.}},
\bauthor{\bsnm{Menotti}, \binits{D.}}:
\bctitle{How the choice of samples for building arrhythmia classifiers impact their performances}.
In: \bbtitle{2011 Annual International Conference of the IEEE Engineering in Medicine and Biology Society},
pp. \bfpage{4988}--\blpage{4991}
(\byear{2011}).
\bcomment{IEEE}
\end{bchapter}
\endbibitem

\bibitem[\protect\citeauthoryear{Luz et~al.}{2016}]{luz2016ecg}
\begin{barticle}
\bauthor{\bsnm{Luz}, \binits{E.J.d.S.}},
\bauthor{\bsnm{Schwartz}, \binits{W.R.}},
\bauthor{\bsnm{C{\'a}mara-Ch{\'a}vez}, \binits{G.}},
\bauthor{\bsnm{Menotti}, \binits{D.}}:
\batitle{Ecg-based heartbeat classification for arrhythmia detection: A survey}.
\bjtitle{Computer methods and programs in biomedicine}
\bvolume{127},
\bfpage{144}--\blpage{164}
(\byear{2016})
\end{barticle}
\endbibitem

\bibitem[\protect\citeauthoryear{Mousavi and Afghah}{2019}]{mousavi2019inter}
\begin{bchapter}
\bauthor{\bsnm{Mousavi}, \binits{S.}},
\bauthor{\bsnm{Afghah}, \binits{F.}}:
\bctitle{Inter-and intra-patient ecg heartbeat classification for arrhythmia detection: a sequence to sequence deep learning approach}.
In: \bbtitle{ICASSP 2019-2019 IEEE International Conference on Acoustics, Speech and Signal Processing (ICASSP)},
pp. \bfpage{1308}--\blpage{1312}
(\byear{2019}).
\bcomment{IEEE}
\end{bchapter}
\endbibitem

\bibitem[\protect\citeauthoryear{Micheli}{2009}]{micheli2009neural}
\begin{barticle}
\bauthor{\bsnm{Micheli}, \binits{A.}}:
\batitle{Neural network for graphs: A contextual constructive approach}.
\bjtitle{IEEE Transactions on Neural Networks}
\bvolume{20}(\bissue{3}),
\bfpage{498}--\blpage{511}
(\byear{2009})
\end{barticle}
\endbibitem

\bibitem[\protect\citeauthoryear{Mathews et~al.}{2018}]{mathews2018novel}
\begin{barticle}
\bauthor{\bsnm{Mathews}, \binits{S.M.}},
\bauthor{\bsnm{Kambhamettu}, \binits{C.}},
\bauthor{\bsnm{Barner}, \binits{K.E.}}:
\batitle{A novel application of deep learning for single-lead ecg classification}.
\bjtitle{Computers in biology and medicine}
\bvolume{99},
\bfpage{53}--\blpage{62}
(\byear{2018})
\end{barticle}
\endbibitem

\bibitem[\protect\citeauthoryear{Mathunjwa et~al.}{2022}]{mathunjwa2022ecg}
\begin{barticle}
\bauthor{\bsnm{Mathunjwa}, \binits{B.M.}},
\bauthor{\bsnm{Lin}, \binits{Y.-T.}},
\bauthor{\bsnm{Lin}, \binits{C.-H.}},
\bauthor{\bsnm{Abbod}, \binits{M.F.}},
\bauthor{\bsnm{Sadrawi}, \binits{M.}},
\bauthor{\bsnm{Shieh}, \binits{J.-S.}}:
\batitle{Ecg recurrence plot-based arrhythmia classification using two-dimensional deep residual cnn features}.
\bjtitle{Sensors}
\bvolume{22}(\bissue{4}),
\bfpage{1660}
(\byear{2022})
\end{barticle}
\endbibitem

\bibitem[\protect\citeauthoryear{Moody and Mark}{1990}]{moody1990bih}
\begin{bchapter}
\bauthor{\bsnm{Moody}, \binits{G.B.}},
\bauthor{\bsnm{Mark}, \binits{R.G.}}:
\bctitle{The mit-bih arrhythmia database on cd-rom and software for use with it}.
In: \bbtitle{[1990] Proceedings Computers in Cardiology},
pp. \bfpage{185}--\blpage{188}
(\byear{1990}).
\bcomment{IEEE}
\end{bchapter}
\endbibitem

\bibitem[\protect\citeauthoryear{Moody and Mark}{2001}]{moody2001impact}
\begin{barticle}
\bauthor{\bsnm{Moody}, \binits{G.B.}},
\bauthor{\bsnm{Mark}, \binits{R.G.}}:
\batitle{The impact of the mit-bih arrhythmia database}.
\bjtitle{IEEE Engineering in Medicine and Biology Magazine}
\bvolume{20}(\bissue{3}),
\bfpage{45}--\blpage{50}
(\byear{2001})
\end{barticle}
\endbibitem

\bibitem[\protect\citeauthoryear{Pandey and Janghel}{2019}]{pandey2019automatic}
\begin{barticle}
\bauthor{\bsnm{Pandey}, \binits{S.K.}},
\bauthor{\bsnm{Janghel}, \binits{R.R.}}:
\batitle{Automatic detection of arrhythmia from imbalanced ecg database using cnn model with smote}.
\bjtitle{Australasian physical \& engineering sciences in medicine}
\bvolume{42}(\bissue{4}),
\bfpage{1129}--\blpage{1139}
(\byear{2019})
\end{barticle}
\endbibitem

\bibitem[\protect\citeauthoryear{Peimankar and Puthusserypady}{2021}]{peimankar2021dens}
\begin{barticle}
\bauthor{\bsnm{Peimankar}, \binits{A.}},
\bauthor{\bsnm{Puthusserypady}, \binits{S.}}:
\batitle{Dens-ecg: A deep learning approach for ecg signal delineation}.
\bjtitle{Expert systems with applications}
\bvolume{165},
\bfpage{113911}
(\byear{2021})
\end{barticle}
\endbibitem

\bibitem[\protect\citeauthoryear{Ren and Jin}{2019}]{ren2019vector}
\begin{barticle}
\bauthor{\bsnm{Ren}, \binits{W.}},
\bauthor{\bsnm{Jin}, \binits{N.}}:
\batitle{Vector visibility graph from multivariate time series: a new method for characterizing nonlinear dynamic behavior in two-phase flow}.
\bjtitle{Nonlinear Dynamics}
\bvolume{97},
\bfpage{2547}--\blpage{2556}
(\byear{2019})
\end{barticle}
\endbibitem

\bibitem[\protect\citeauthoryear{Shoughi and Dowlatshahi}{2021}]{shoughi2021practical}
\begin{bchapter}
\bauthor{\bsnm{Shoughi}, \binits{A.}},
\bauthor{\bsnm{Dowlatshahi}, \binits{M.B.}}:
\bctitle{A practical system based on cnn-blstm network for accurate classification of ecg heartbeats of mit-bih imbalanced dataset}.
In: \bbtitle{2021 26th International Computer Conference, Computer Society of Iran (CSICC)},
pp. \bfpage{1}--\blpage{6}
(\byear{2021}).
\bcomment{IEEE}
\end{bchapter}
\endbibitem

\bibitem[\protect\citeauthoryear{Scarselli et~al.}{2008}]{scarselli2008graph}
\begin{barticle}
\bauthor{\bsnm{Scarselli}, \binits{F.}},
\bauthor{\bsnm{Gori}, \binits{M.}},
\bauthor{\bsnm{Tsoi}, \binits{A.C.}},
\bauthor{\bsnm{Hagenbuchner}, \binits{M.}},
\bauthor{\bsnm{Monfardini}, \binits{G.}}:
\batitle{The graph neural network model}.
\bjtitle{IEEE transactions on neural networks}
\bvolume{20}(\bissue{1}),
\bfpage{61}--\blpage{80}
(\byear{2008})
\end{barticle}
\endbibitem

\bibitem[\protect\citeauthoryear{Sraitih et~al.}{2021}]{sraitih2021overview}
\begin{bchapter}
\bauthor{\bsnm{Sraitih}, \binits{M.}},
\bauthor{\bsnm{Jabrane}, \binits{Y.}},
\bauthor{\bsnm{Atlas}, \binits{A.}}:
\bctitle{An overview on intra-and inter-patient paradigm for ecg heartbeat arrhythmia classification}.
In: \bbtitle{2021 International Conference on Digital Age \& Technological Advances for Sustainable Development (ICDATA)},
pp. \bfpage{1}--\blpage{7}
(\byear{2021}).
\bcomment{IEEE}
\end{bchapter}
\endbibitem

\bibitem[\protect\citeauthoryear{Shuman et~al.}{2013}]{shuman2013emerging}
\begin{barticle}
\bauthor{\bsnm{Shuman}, \binits{D.I.}},
\bauthor{\bsnm{Narang}, \binits{S.K.}},
\bauthor{\bsnm{Frossard}, \binits{P.}},
\bauthor{\bsnm{Ortega}, \binits{A.}},
\bauthor{\bsnm{Vandergheynst}, \binits{P.}}:
\batitle{The emerging field of signal processing on graphs: Extending high-dimensional data analysis to networks and other irregular domains}.
\bjtitle{IEEE signal processing magazine}
\bvolume{30}(\bissue{3}),
\bfpage{83}--\blpage{98}
(\byear{2013})
\end{barticle}
\endbibitem

\bibitem[\protect\citeauthoryear{Sperduti and Starita}{1997}]{sperduti1997supervised}
\begin{barticle}
\bauthor{\bsnm{Sperduti}, \binits{A.}},
\bauthor{\bsnm{Starita}, \binits{A.}}:
\batitle{Supervised neural networks for the classification of structures}.
\bjtitle{IEEE Transactions on Neural Networks}
\bvolume{8}(\bissue{3}),
\bfpage{714}--\blpage{735}
(\byear{1997})
\end{barticle}
\endbibitem

\bibitem[\protect\citeauthoryear{Sun et~al.}{2014}]{sun2014characterizing}
\begin{barticle}
\bauthor{\bsnm{Sun}, \binits{X.}},
\bauthor{\bsnm{Small}, \binits{M.}},
\bauthor{\bsnm{Zhao}, \binits{Y.}},
\bauthor{\bsnm{Xue}, \binits{X.}}:
\batitle{Characterizing system dynamics with a weighted and directed network constructed from time series data}.
\bjtitle{Chaos: An Interdisciplinary Journal of Nonlinear Science}
\bvolume{24}(\bissue{2}),
\bfpage{024402}
(\byear{2014})
\end{barticle}
\endbibitem

\bibitem[\protect\citeauthoryear{Shobanadevi and Veeramakali}{2023}]{shobanadevi2023classification}
\begin{botherref}
\oauthor{\bsnm{Shobanadevi}, \binits{A.}},
\oauthor{\bsnm{Veeramakali}, \binits{T.}}:
Classification and interpretation of ecg arrhythmia through deep learning techniques
(2023)
\end{botherref}
\endbibitem

\bibitem[\protect\citeauthoryear{Wu et~al.}{2020}]{wu2020comprehensive}
\begin{barticle}
\bauthor{\bsnm{Wu}, \binits{Z.}},
\bauthor{\bsnm{Pan}, \binits{S.}},
\bauthor{\bsnm{Chen}, \binits{F.}},
\bauthor{\bsnm{Long}, \binits{G.}},
\bauthor{\bsnm{Zhang}, \binits{C.}},
\bauthor{\bsnm{Philip}, \binits{S.Y.}}:
\batitle{A comprehensive survey on graph neural networks}.
\bjtitle{IEEE transactions on neural networks and learning systems}
\bvolume{32}(\bissue{1}),
\bfpage{4}--\blpage{24}
(\byear{2020})
\end{barticle}
\endbibitem

\bibitem[\protect\citeauthoryear{Xu et~al.}{2018}]{xu2018powerful}
\begin{botherref}
\oauthor{\bsnm{Xu}, \binits{K.}},
\oauthor{\bsnm{Hu}, \binits{W.}},
\oauthor{\bsnm{Leskovec}, \binits{J.}},
\oauthor{\bsnm{Jegelka}, \binits{S.}}:
How powerful are graph neural networks?
arXiv preprint arXiv:1810.00826
(2018)
\end{botherref}
\endbibitem

\bibitem[\protect\citeauthoryear{Ye et~al.}{2012}]{ye2012heartbeat}
\begin{barticle}
\bauthor{\bsnm{Ye}, \binits{C.}},
\bauthor{\bsnm{Kumar}, \binits{B.V.}},
\bauthor{\bsnm{Coimbra}, \binits{M.T.}}:
\batitle{Heartbeat classification using morphological and dynamic features of ecg signals}.
\bjtitle{IEEE Transactions on Biomedical Engineering}
\bvolume{59}(\bissue{10}),
\bfpage{2930}--\blpage{2941}
(\byear{2012})
\end{barticle}
\endbibitem

\bibitem[\protect\citeauthoryear{Zhou et~al.}{2020}]{zhou2020graph}
\begin{barticle}
\bauthor{\bsnm{Zhou}, \binits{J.}},
\bauthor{\bsnm{Cui}, \binits{G.}},
\bauthor{\bsnm{Hu}, \binits{S.}},
\bauthor{\bsnm{Zhang}, \binits{Z.}},
\bauthor{\bsnm{Yang}, \binits{C.}},
\bauthor{\bsnm{Liu}, \binits{Z.}},
\bauthor{\bsnm{Wang}, \binits{L.}},
\bauthor{\bsnm{Li}, \binits{C.}},
\bauthor{\bsnm{Sun}, \binits{M.}}:
\batitle{Graph neural networks: A review of methods and applications}.
\bjtitle{AI Open}
\bvolume{1},
\bfpage{57}--\blpage{81}
(\byear{2020})
\end{barticle}
\endbibitem

\bibitem[\protect\citeauthoryear{Zhao et~al.}{2022}]{zhao2022ecgnn}
\begin{bchapter}
\bauthor{\bsnm{Zhao}, \binits{X.}},
\bauthor{\bsnm{Liu}, \binits{Z.}},
\bauthor{\bsnm{Han}, \binits{L.}},
\bauthor{\bsnm{Peng}, \binits{S.}}:
\bctitle{Ecgnn: Enhancing abnormal recognition in 12-lead ecg with graph neural network}.
In: \bbtitle{2022 IEEE International Conference on Bioinformatics and Biomedicine (BIBM)},
pp. \bfpage{1411}--\blpage{1416}
(\byear{2022}).
\bcomment{IEEE}
\end{bchapter}
\endbibitem

\bibitem[\protect\citeauthoryear{Zaor{\'a}lek et~al.}{2018}]{zaoralek2018patient}
\begin{barticle}
\bauthor{\bsnm{Zaor{\'a}lek}, \binits{L.}},
\bauthor{\bsnm{Plato{\v{s}}}, \binits{J.}},
\bauthor{\bsnm{Sn{\'a}{\v{s}}el}, \binits{V.}}:
\batitle{Patient-adapted and inter-patient ecg classification using neural network and gradient boosting}.
\bjtitle{Neural Network World}
\bvolume{28}(\bissue{3}),
\bfpage{241}--\blpage{254}
(\byear{2018})
\end{barticle}
\endbibitem

\bibitem[\protect\citeauthoryear{Zhang and Small}{2006}]{zhang2006complex}
\begin{barticle}
\bauthor{\bsnm{Zhang}, \binits{J.}},
\bauthor{\bsnm{Small}, \binits{M.}}:
\batitle{Complex network from pseudoperiodic time series: Topology versus dynamics}.
\bjtitle{Physical review letters}
\bvolume{96}(\bissue{23}),
\bfpage{238701}
(\byear{2006})
\end{barticle}
\endbibitem

\bibitem[\protect\citeauthoryear{Zhao and Zhang}{2005}]{zhao2005ecg}
\begin{bchapter}
\bauthor{\bsnm{Zhao}, \binits{Q.}},
\bauthor{\bsnm{Zhang}, \binits{L.}}:
\bctitle{Ecg feature extraction and classification using wavelet transform and support vector machines}.
In: \bbtitle{2005 International Conference on Neural Networks and Brain},
vol. \bseriesno{2},
pp. \bfpage{1089}--\blpage{1092}
(\byear{2005}).
\bcomment{IEEE}
\end{bchapter}
\endbibitem

\end{thebibliography}

\end{document}